\documentclass[%
 reprint,
superscriptaddress,
unsortedaddress,
 amsmath,amssymb,
onecolumn,
]{revtex4-2}

\usepackage{graphicx}% Include figure files
\usepackage{dcolumn}% Align table columns on decimal point
\usepackage{bm}% bold math
\usepackage{hyperref}% add hypertext capabilities
\usepackage{xcolor}

\begin{document}

\preprint{}

\author{Marco Baldovin}
 \affiliation{Dipartimento di Fisica, Universit\`a ``Sapienza'', Rome  I-00185, Italy}%Lines break automatically or can be forced with \\
\affiliation{Universit\'{e}  Paris-Saclay,  CNRS,  LPTMS,  91405,  Orsay,  France}%
 \email{marco.baldovin@universite-paris-saclay.fr}
\author{Fabio Cecconi}
\affiliation{CNR-Istituto dei Sistemi Complessi and INFN, Unità di Roma 1, I-00185, Rome, Italy}%
\author{Antonello Provenzale}%
\affiliation{CNR-Istituto di Geoscienze e Georisorse, Pisa, I-56124 Italy}%
\author{Angelo Vulpiani}%
\affiliation{Dipartimento di Fisica, Universit\`a ``Sapienza'', Rome  I-00185, Italy}%

\date{September 12, 2022}% It is always \today, today,

  \newcommand{\av}[1]{\left\langle#1\right\rangle}
  \newcommand{\cbr}[1]{\left(#1\right)}
  \newcommand{\sbr}[1]{\left[#1\right]}
  \newcommand{\bbr}[1]{\left\{#1\right\}}
  
  \newcommand{\moby}{\mu}
  \newcommand{\frict}{\gamma}
  \newcommand{\mean}{\lambda}
  \newcommand{\tmean}{\widetilde{\lambda}}
  \newcommand{\stiff}{k}
  \newcommand{\tstiff}{\widetilde{k}}
  \newcommand{\stud}{k}
  \newcommand{\tstud}{\widetilde{k}}
  
  \newcommand{\dimless}[1]{{#1}^*}

  \newcommand{\addnew}[1]{#1}

\title{Extracting causation from millennial-scale climate fluctuations in the last 800 kyr}

\DeclareRobustCommand{\vect}[1]{
  \ifcat#1\relax
    \boldsymbol{#1}
  \else
    \mathbf{#1}
  \fi}

\newcommand{\newversion}[1]{#1}

%\affil[+]{these authors contributed equally to this work}

%\keywords{Keyword1, Keyword2, Keyword3}

\begin{abstract}
The detection of cause-effect relationships from the analysis of paleoclimatic records is a crucial step to disentangle the main mechanisms at work in the climate system. Here, we { show that} the approach based on the generalized Fluctuation-Dissipation Relation, complemented by the analysis of the Transfer Entropy, { allows the causal links} to be identified between temperature, CO$_2$ concentration and astronomical
forcing during the glacial cycles of the last 800 kyr based on 
Antarctic ice core records. 
When considering the whole spectrum of time scales, the 
results of the analysis
suggest that temperature drives CO$_2$ concentration, or that are both 
driven by the
common astronomical forcing. However, considering only
millennial-scale fluctuations, the results reveal { the presence of more complex} causal links, indicating
that CO$_2$ variations { contribute to driving} the changes of temperature on such time scales. { The results also evidence a slow temporal variability in the strength of the millennial-scale causal links between temperature and CO$_2$ concentration.}
\end{abstract}

\flushbottom
\maketitle
\thispagestyle{empty}

\noindent 

%%%%%%%%%%%%%%%%%%%%%%%%%%%%%%%%%%%%%%%%%%%%%%%%%%%%%%%%%%%%%
\section*{Introduction} \label{sec:introduction}
%%%%%%%%%%%%%%%%%%%%%%%%%%%%%%%%%%%%%%%%%%%%%%%%%%%%%%%%%%%%%
Earth’s climate is a complex nonlinear system in which  multiple feedback mechanisms control the stability, variability and/or abrupt transitions between different climatic states (see e.g. Refs.\cite{Pierrehumbert10,ghilluca}). Such feedbacks generate internal, intrinsic climatic oscillations and can amplify or damp the effects of external forcing factors, see e.g. Refs.\cite{stoc,Ghil94,enso2,enso1,Ghil,luca}. 
In such a framework, identifying cause-effect relationships from an observed behavior is often difficult, and refined mathematical approaches and data 
analysis methods able to go beyond correlation estimates are needed to disentangle causation links, as discussed for example in Refs.\cite{RungeNatComm19,runge2014quantifying,moon2017unified}.

One outstanding example of climate variability is the case of glacial-interglacial oscillations of the Pleistocene. In the last three million years, Earth’s climate fluctuated between prolonged glacial periods, slowly developing through global temperature decrease and the build-up of extended ice sheets, and shorter interglacials with milder climate, generated by a relatively rapid (in geological sense) melting of the ice\cite{Hays,Imbrie93,Lisieki,Raymo}.  Such glacial cycles are believed to be a nonlinear reaction of the climate, or of some of its sub-systems, to the slow variation of the orbital forcing via amplifying feedbacks \cite{milanko41}. 
 The Antarctic ice cores drilled at Vostok\cite{Petit} and, more recently, by the EPICA project\cite{EPICA} have revealed the details of the glacial oscillations in the last few hundred thousand years\cite{Luthi08,Bereiter15}. Approximately synchronous variations of the reconstructed Antarctic temperature and of carbon dioxide concentration in the paleo-atmosphere are visible 
 \cite{siegenthaler,Parrenin}, { although the precise lead-lag relationships between (Antarctic) temperature and CO$_2$ concentration are still a matter of debate. In particular, such lead-lag relations could vary on both the time scale and the specific period considered}\cite{vanNes}. { For example,} a detailed correlation analysis of { a high-resolution Antarctic ice core has indicated that during the last glacial period there is a lagged variation of CO$_2$ with respect to temperature on millennial time scales, { which however becomes more complex at centennial time scales} \cite{Bauska}. { { The interpretation of such time-scale dependence of the lead-lag relationships between CO$_2$ and $T$ can be offered in terms of the presence of multiple mechanisms at different time scales}} \cite{vanNes}. { One possibility is the interplay of a slower process associated with the reorganization of the Southern Ocean carbon cycle, and faster (possibly abrupt) processes associated with Northern Hemisphere Dansgaard-Oeschger events}}\cite{DO,Bauska}. Of course, whether the dynamics have been
considered synchronous depends a lot on the quality of the age model and on whether lagged correlation or actual differences in specific change points are considered.

On the other hand, the
correlation between two variables is not, in itself, a reliable measure of causation, { as already pointed out in Ref.\cite{vanNes} for paleoclimate dynamics}. 
A typical case in which correlation fails to catch the underlying causal structure is when two mutually independent variables $x_1(\tau)$ and $x_2(\tau)$ are driven by a common forcing $f(\tau)$, $\tau$ being the absolute time. In this case, a strong correlation can be easily misinterpreted as causation. This situation is frequently encountered in the climate system, as discussed in Ref.\cite{runge2014quantifying}.
A relevant challenge is thus to disentangle the cause-effect relationships from the analysis of the two signals. 
In past works on glacial oscillations, the issue of causation has been addressed by the work in Ref.\cite{stips2016causal}, which looked at the causal structure of the temperature-CO$_2$ concentration relations using an information flow approach \cite{shreiber00}. 
However, confounding factors can be present and different causal relationships can exist on different time scales\cite{vanNes}. These issues should not be overlooked and they are further addressed in the present manuscript.

One reliable definition of cause-effect relationship, able to take into account different behaviors at different scales, is based on the observation of the average trajectory of $x_2(\tau)$ after an active perturbation of the variable $x_1(\tau)$ has been performed. This kind of ``probing'' resembles the idea behind the mathematical formal definition of causation by Judea Pearl~\cite{PearlBook}. In dynamical systems it can be characterized by looking at the linear response matrix function~\cite{aurell2016causal, baldovin20}
\begin{equation}
 \label{eq:resp_obs}
R_{ij}(t;\tau) = \frac{\overline{\delta x_i(\tau+t)}}{\delta x_j(\tau)}.
\end{equation}
Here $\delta x_j(\tau)$ is the value of an instantaneous, external perturbation operated on the variable $x_j$ at time $\tau$, while $\overline{\delta x_i(\tau+t)}$ is the average (over many repetitions of the experiment) of the difference between the perturbed trajectory of $x_i(\tau)$ and its unperturbed evolution.  In what follows, we will assume that the above defined quantity does not depend on the absolute time $\tau$, but only on the lag $t$, so that $R_{ij}(t;\tau)\equiv R_{i,j}(t)$. From Eq.~\eqref{eq:resp_obs} it can be deduced that ${R}_{ij}(0)=\delta_{ij}$ by definition, where $\delta_{ij}$ is the Kronecker-delta. Thus the diagonal entries decay from the starting value 
$R_{ii}(0) = 1$, while the off-diagonal entries grow from the starting value $R_{i\ne j} = 0$. From a physical point of view, this simply means that the variables of a system cannot generate an immediate reaction on the others, as some (possibly very small) delay has to occur between a ``cause'' and its ``effect''.

Of course, \eqref{eq:resp_obs} cannot be applied to any problem for which only time series referring to past events are available. 
%it is obviously not possible to perturb events that have already occurred. 
However, a series of well known results of response theory show that $R_{ij}(t)$ can be written in terms of time correlations of suitable quantities~\cite{falcioni90,baldovin21}. One of the possible formulations of this principle, sometimes called generalized Fluctuation-Dissipation Relation (generalized FDR), is:
\begin{equation}
 \label{eq:resp_log}
 R_{ij}(t) = -\left\langle x_i(t)\frac{\partial}{\partial x_j}\log P(\vect{x})\big|_{\vect{x}(0)}\right\rangle.
\end{equation}
Here $P(\vect{x})$ is the stationary probability distribution of the whole phase-space vector $\vect{x}=(x_1,x_2, ..., x_n)$ describing the system dynamics; the vector $\vect{x}$ is meant to include all the variables that determine the behavior of $x_i$ and $x_j$. 

In particular, if the dynamics of a system of $n$ variables $x_1,\dots,x_n$ is linear, the matrix of the response functions simplifies to (see Supplemental Material for a complete derivation)
\begin{equation}
\label{eq:respcorr}
\mathbb{R}(t)=\mathbb{C}(t) \mathbb{C}^{-1}(0)\,,
\end{equation}
that is easily determined from the elements of the correlation matrix, $C_{ij}(t)= \langle x_i(\tau + t)x_j(\tau)\rangle$, of the available data sets [where $\mathbb{C}^{-1}(0)$ is the inverse matrix of $C_{ij}(0) = \langle x_i(\tau)x_j(\tau)\rangle$].
{Hereafter, we will denote with $\mathbb{M}$ a matrix and with $M_{ij}$ the scalar values of its entries}.
{To appreciate the difference with a simple correlations analysis, it is} {useful} { to explicitly write the matrix \eqref{eq:respcorr} for a two-dimensional system $[x(t),y(t)]$, where the inversion of $\mathbb{C}(0)$ can be easily performed.} {We assume that $x(t)$ and $y(t)$ have zero average and unitary variance, as in the following we will always consider normalized signals of this sort.}
{ In this simple case, the 2$\times$2 response matrix reads \begin{equation}
\label{eq:rmatrix}
\mathbb{R}(t) = 
\begin{pmatrix}
\dfrac{C_{xx}(t) - C_{xy}(0)\;C_{xy}(t)}{1-C_{xy}^2(0)} & 
\dfrac{C_{xy}(t) - C_{xy}(0)\;C_{xx}(t)}{1-C_{xy}^2(0)} 
\\
& \\ 
\dfrac{C_{yx}(t) - C_{xy}(0)\;C_{yy}(t)}{1-C_{xy}^2(0)} & 
\dfrac{C_{yy}(t) - C_{xy}(0)\;C_{yx}(t)}{1-C_{xy}^2(0)} 
\end{pmatrix}.
\end{equation}}
{The relation $C_{xx}(0) = C_{yy}(0) = 1$, following from the normalization of the data, has been used. In general, $C_{xy}(t)$ is different from $C_{yx}(t)$, the symmetry $C_{xy}(0) = C_{yx}(0)$ holding true only for $t=0$.}
We remark that, even if the response can be computed as a combination of correlation functions, it provides information about the causal structure of the system which could not be deduced from cross-correlations alone~\cite{aurell2016causal}. 
{Note that this result is quite robust: the presence of small nonlinearities in the dynamics is not expected to spoil the ability of Eq.~\eqref{eq:respcorr} to detect causal links~\cite{baldovin20}.}

In principle, the generalized FDR solves the problem of inferring causal relations for any dynamical system, but its application strongly relies on the assumption that the dynamics of the chosen set of observables does not depend on any variable that is external to the system (i.e., in the language of stochastic processes, that the dynamics is Markovian). Moreover, the shape of the steady state distribution has to be known, at least approximately.

The application of this kind of analysis to the EPICA paleoclimate data is thus limited by two factors: (i) the lack of knowledge of a proper set of variables $\vect{x}=(x_1, ... , x_n)$ fully describing a Markovian dynamics and
(ii) the relative shortage of data (about 1600 measurements, covering a range of 800 kyr), which would not allow a reliable estimate of the joint probability distribution $P(\vect{x})$, even if a valid set of observables $\vect{x}$ was known.

In this paper, we focus on the variations of the carbon dioxide concentration, [CO$_2$], and the reconstructed temperature $T$,  during glacial-interglacial oscillations.
{ Assuming a sufficient time-scale separation between the  astronomical forcing (with a typical time of the order of $20$ kyr or more) and the internal climate variability on millennial or shorter time scales, we} 
delineate a qualitative analysis of the causation relations between [CO$_2$] and $T$ in the last 800 kyr. The strategy is based on the definition of a proper set of ``fast'' components whose dynamics 
{ is assumed to}
 mainly reflect internal climate variability, associated with the interaction of the different climate sub-systems. Notice, however, that the intensity of such fast fluctuations could be modulated by the climate background state and thus by the astronomical forcing\cite{kawa}. { Here, the important hypothesis is that the fast oscillations are not completely ``slaved" to the slow forcing.} Within the reasonable assumption that the mutual interactions of the fast components can be approximately linearized,
the response on fast time-scale of the order of one to two kyr can be inferred by exploiting Eq.~\eqref{eq:respcorr}, which, for these variables, can be evaluated with good accuracy even with the available quantity of data.  Special attention should be given to the temporal resolution of the paleoclimatic data, which, in some portions of the record, may hamper the ability to safely detect millennial-scale oscillations, a point that is further addressed below and in the Methods section. { Important messages of this work are that (a) FDR analysis provides relevant information on the causation relationships in (paleo)climate signals, (b) considering only unfiltered data including all time scales can lead to incomplete results, masking the possible emergence of more complex causal relationships on specific time scales, { and (c) there is a clear long-term temporal variation in the strength of the millennial-scale causal relationships between temperature and CO$_2$ concentration.}}

%%%%%%%%%%%%%%%%%%%%%%%%%%%%%%%%%%%%%%%%%%%%%%%%%%%%%%%%%%%%%%%%%%%%
\section*{Results}
%%%%%%%%%%%%%%%%%%%%%%%%%%%%%%%%%%%%%%%%%%%%%%%%%%%%%%%%%%%%%%%%%%%%
As discussed in the Introduction, the generalized FDR can be used to unravel causal links between the variables of a physical system by analyzing their correlations, provided that (i) the dynamics is not subject to an external common driving which simultaneously forces several variables and (ii) the stationary distribution is known, at least approximately. The coupled dynamics of $T$ and [CO$_2$] in the last 800 kyr does not fulfill any of these two conditions, as their behaviour is heavily conditioned by the external astronomical driving, and the relatively small amount of available data (about 1600 measurements spanning the whole record) does not allow to reconstruct a reliable coupled probability distribution for the two quantities.
%----------------------------------------------------------------------
\begin{figure}[]
\centering
  \includegraphics[width=\linewidth]{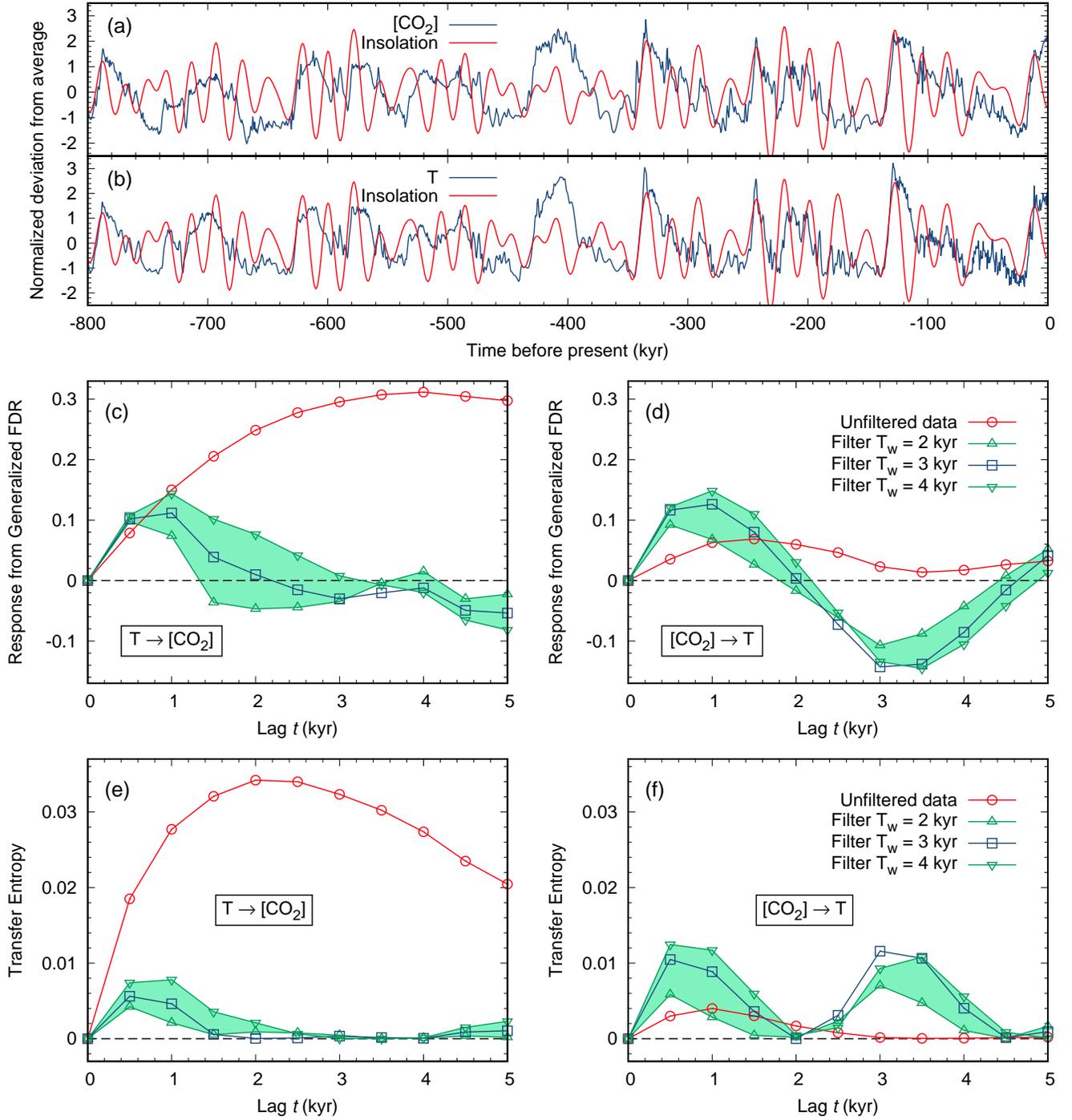}
  \caption{\label{fig:fdrclimate} 
  Analysis of the mutual influence between 
  temperature $T$ and CO$_2$ concentration in paleoclimate data.
  Panels (a) and (b) show the deviation from average of the two signals as a function of time (blue curves), normalized by the standard deviation. The daily mean insolation at $65^{\circ}$ N summer solstice}, revealing the typical time scales of the external driving, is also reported (red curves; see Methods sections for details on the data sources).
  In panels (c) and (d) the response function, computed according to the Generalized FDR, is plotted. {The analytical formula is given by the non-diagonal elements $R_{xy}(t)$ and $R_{yx}(t)$ of the linear response matrix~\eqref{eq:rmatrix}, where $x$ and $y $ are the normalized [CO$_2$] and $T$ signals shown in panels (a), (b) (and their high-pass filtered analogues).} Panel (c) refers to the effect of $T$ on [CO$_2$], while panel (d) shows the opposite relation.  Red circles represent the results of a direct application of Eq.~\eqref{eq:respcorr} on raw data, apparently suggesting that $T\to$[CO$_2$] is stronger than [CO$_2$]$\to T$. 
  The response on data filtered over $T_w=3$ kyr window (blue squares) instead indicates that the impact of [CO$_2$] on $T$ becomes larger. % and it vanishes within about $2$ kyr.
  The result is robust with respect of  $T_w$ variations by one kyr (green up/down triangles). A similar analysis, where TE are computed instead of generalised FDR, is shown in Panels (e) and (f). { Here, we have considered the data for which the temporal resolution of the temperature record has been degraded to become similar to that of CO$_2$ concentration, as discussed in the Methods section. For the undegraded temperature data, the role of CO$_2$ driving is even larger, see 
  {Fig.~S5 of} the Supplemental Material.} 
\end{figure}
%----------------------------------------------------------------------

In the Methods section, we show that both the above difficulties can be circumvented as long as there is a { sufficient} scale separation between the (slow) typical time scale of the external driving and the (short) characteristic times of the interaction between the variables. 
This is indeed the case for Pleistocene climatic variability, where the  time scale of the external astronomical forcing is of the order of 20 kyr or more (Milankovitch cycles \cite{milanko41}), while intense climatic fluctuations occur on a much faster time scale, of about 1 kyr.

The key ingredient for the analysis is thus a proper high-pass temporal filtering of the signals: before applying the FDR, we subtract from the time series of $T$ and [CO$_2$] a running average (see Methods for details) over windows, $T_w$, of a few kyr. 
This basically allows to filter out any possible spurious correlation due for example to a common influence of the slow external forcing, while keeping the relevant information on the short-time mutual relationship. 
In addition, the distribution of the filtered variables is {approximately} Gaussian (Methods, Fig~\ref{fig:distrclimate}): this is consistent with the working hypothesis that the dynamics on the fast scales is approximately linear, and that the use of FDR in the form of Eq.~\eqref{eq:respcorr} is justified.

The results of the analysis are shown in Fig.~\ref{fig:fdrclimate}, where the  response function computed with the generalized FDR is plotted as a function of time.  The 
approach based on a 
straightforward application of Eq.~\eqref{eq:respcorr} to the 
unfiltered data would suggest that the relative influence of temperature on CO$_2$ concentration, $T\rightarrow [$CO$_2]$, is { much} stronger than the reversed causal link $[$CO$_2]\rightarrow T$. 
When the formula is instead applied to the filtered data, a 
different scenario is observed. { For the whole time series,}
the relative influence of the temperature on [CO$_2$] is at most $\simeq 0.1$ (in a scale in which the self-response at time 0 is set equal to 1), and it almost vanishes { for lags beyond 2 kyr}; on the other hand the intensity of the $[$CO$_2]\rightarrow T$ causal relationship 
%dramatically 
increases with respect to 
what is observed without high-pass filtering, doubling that of the reverse relation on the scale of 1 kyr. { The causal link $[$CO$_2]\rightarrow T$ does now vanish at longer lags, displaying an oscillating behavior as shown also by the Transfer Entropy results.}

As discussed in the Methods section, the width $T_w$ of the time window for the filter has been chosen to be 3 kyr.
Figure~\ref{fig:fdrclimate} also shows that the method is quite robust with respect to the choice of $T_w$: indeed, we obtained an analogous behaviour of the response functions using $T_w = 2$ kyr and $T_w = 4$ kyr. 

{  An important point concerns the temporal resolution of the two signals considered. {Figure~S4 of the Supplemental Material} shows that the resolution of the CO$_2$ record is generally coarser than that of temperature, and both vary in time. To avoid possible spurious effects generated by the different temporal resolution and the different weight of the spline interpolation, we opted for degrading the resolution of the temperature signal to that of the carbon dioxide concentration {(and vice-versa, in the few intervals where the temporal resolution of [CO$_2$] is larger than that of $T$)}. 
Considering instead the results of the original (non degraded) data, as shown in Fig.~S5 of the Supplemental Material, the main messages do not change.
}

A similar analysis, in which the transfer entropy (TE) between the two variables is computed before and after the filtering procedure [Figg.~\ref{fig:fdrclimate}(e) and~\ref{fig:fdrclimate}(f) ], confirms that a high-pass 
filtering of the data is crucial to elucidate the causal relationships between temperature and CO$_2$ concentration in the last 800 kyr of the Pleistocene. A brief discussion on the concept of TE, which may be regarded as complementary to that of response, is given in the Methods section, where also further details about its application in this context are provided. Here it is worth noticing that the qualitative  results are similar to those achieved with the generalized FDR approach: a 
direct application of the formula to the raw data would suggests a large influence of $T$ on [CO$_2$]; on the other hand, a proper temporal filtering { a more complex picture}. A remarkable similarity between the behaviour of the two observables can be appreciated in Panels~\ref{fig:fdrclimate}(d) and~\ref{fig:fdrclimate}(f) (recalling that TE, at variance with response, is a positive-definite quantity).

%%%%%%%%%%%%%%%%%%%%%%%%%%%%%%%%%%%%%%%%%%%%%%%%%%%%%%%%%%%%%%%%
\section*{Discussion}
%%%%%%%%%%%%%%%%%%%%%%%%%%%%%%%%%%%%%%%%%%%%%%%%%%%%%%%%%%%%%%%%
The issue of which climatic signal drives which in the glacial-interglacial record is widely debated. For example, the analysis of Caillon et al.\cite{Caillon} indicated that CO$_2$ lagged Antarctic deglacial warming by 800 $\pm$ 200 years during a specific deglaciation event (Termination III ~ 240,000 years ago). Subsequently, Parrenin et al.\cite{Parrenin} found no asinchronicity between Antarctic temperatures and CO$_2$ variations in the last deglaciation event (Termination TI), even though the situation is not always clear and it can vary with time
%during a deglaciation event
\cite{Pedro, Landais}. The work of Stips et al.\cite{stips2016causal}, based on the use of the information flow to detect causal relations, revealed a complex pattern of cause-effect links, with a predominance of the Antarctic temperature driving CO$_2$ concentration when the whole record is considered. The analysis of millennial-scale fluctuations in the last glacial period showed { that CO$_2$ seems to lag temperature by 500-1000 yr \cite{Bauska}, while more complex relationships may exist on centennial time scales}. Finally, the work of van Nes et al.\cite{vanNes} concluded that different relationships can exist on different time scales. Clearly, crucial to all these lag analyses is the availability of a safely calibrated time scale for both temperature and CO$_2$. In any case, here we found that the maximum value of the FDR for the effect of the CO$_2$ concentration on temperature for the high-pass filtered data is found between 500 and 1500 yr when considering the whole 800-kyr record. Thus, even an uncertainty of the age model of the order of 500 yr does not qualitatively change the results.

Here, we analysed causality links adopting the generalized Fluctuation-Dissipation Relation (see the Methods section for a detailed discussion of how this approach works), further confirming the results using the Transfer Entropy method. The main finding is that, using the data from Refs.\cite{Luthi08,Bereiter15}, we detect a causal link of temperature on [CO$_2$] when considering the whole unfiltered record that includes both millennial-scale fluctuations and longer-term glacial-interglacial oscillations, in keeping with previous results\cite{stips2016causal}. 
On the scales of the astronomical forcing, albedo changes could drive temperature variations and consequently affect the whole cascade of climatic processes, including CO$_2$ changes. The causal link T $\rightarrow$ [CO$_2$] could thus be generated either by slow climatic processes, such as the global ocean's temperature-dependent ability to store CO$_2$, or simply reflect the fact that both climatic signals are controlled by a common driver, namely, the astronomical forcing with the related changes in summer solar insolation at high latitudes.

%----------------------------FIG.2-------------------------------------
\begin{figure}[]
\centering
  \includegraphics[width=\linewidth]{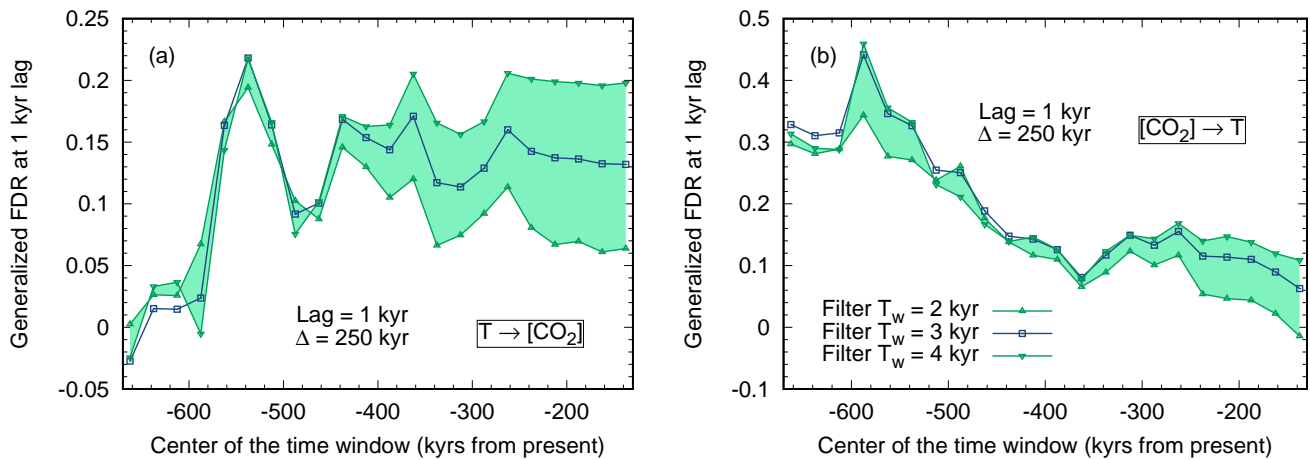}
  \caption{\label{fig:temporal} 
 { Values of the Generalized FDR at lag $t=1$ kyr for the link $T \to$ [CO$_2$] (a) and [CO$_2$] $\to T$ (b), computed in a moving window with width $\Delta=250$ kyr, slided along the whole record. All  details as in panels (c) and (d) of Fig.1. }
 }
\end{figure}
%----------------------------------------------------------------------

On the other hand, the significant novelty of our analysis is that the high-pass filtered paleoclimatic signal, including only fluctuations on scales of a few millennia, displays { a more complex pattern of causal relationships, with mutual driving of [CO$_2$] and temperature. 
%and  a slightly stronger direct causal link of the CO$_2$ concentration on temperature. 
}
This result is robust with respect to the precise value of the threshold used in the high-pass filter, which was varied between 2 and 4 kyr.

{ The results shown in Fig.1 refer to the whole 800-kyr temporal period covered by the record, and they differ from the outcomes of some of previous analyses, performed on other signals spanning a more limited time range\cite{Bauska}.} 
 { This supports the view that the strength of the causal links between temperature and CO$_2$, or more generally between the various components of the climate system, can vary with time, in line with the conclusions of Ref.\cite{vanNes}. Such view is confirmed by Figs.~S6 and~S7 of the Supplemental Material, where we analysed the first or the second half  
 of the record. The results indicated that the strength of the causal links is different in the two periods. On millennial time scales, we always observed a mutual effect between [CO$_2$] and $T$. In particular, the link [CO$_2$] $\to T$ is much stronger in the first 400 kyr of the record  (Fig.~S6), while in the last 400 kyr the reverse link $T \to$ [CO$_2$] becomes more relevant (Fig.~S7). }
 
{ To further explore this issue, in Fig.~\ref{fig:temporal} we show the value of the Generalized FDR for $T \to$ [CO$_2$] and [CO$_2$] $\to T$ at lag 1 kyr (where the first maximum of the FDR is generally located), for a moving window of 250 ky slided along the whole time series. Interestingly, the effect of [CO$_2$] on temperature is stronger in the earlier part of the record and it decreases in the course of time, while the reverse link $T \to$ [CO$_2$] grows in the early part of the record and then stabilizes at an approximately stationary value. Towards the end of the record, the two reverse links have comparable strength. At present, given the limited amount of data it can be difficult to disentangle between real variability and statistical fluctuations, but the results suggest a variability in the relative importance of the causal links between temperature and CO$_2$ concentration. { As a word of caution, we also mention that some of the inferred changes in causal relationships could be a result of changes in the data properties and
the possibly non-stationary resolution properties of the two time series.}
In any case, these complex causal relationships would have been completely lost if we had considered only the unfiltered data.} 
 
 A detailed analysis of the climatic processes inducing millennial-scale changes in CO$_2$ concentration is beyond the scope of this work. Here, we simply mention that the causes of the fluctuations of CO$_2$ concentration on such time scales are widely debated and still not fully clear, but most interpretations involve the role of CO$_2$ outgassing associated with changes in the ocean overturning circulation and/or marine ecosystem functioning\cite{gott,shin}. The negative value of the FDR, observed at a lag of about 3 kyr, can also point to a coupled oscillation in the climate system, although its nature remains currently undetermined. In any case, the results of our analysis support the view that internal climate mechanisms, rather than direct orbital forcing, are responsible for the main variability at millennial time scales in the last 800 kyr, in keeping with the conclusions of Ref.\cite{vanNes}. Further work using simple models such as that of Ref.\cite{Bauska} could help further addressing this issue; { in this respect, it is worth noticing that the conclusion of Ref.\cite{Bauska} are consistent with our results for the last part of the analyzed time interval (see Fig.~\ref{fig:temporal}).}
 
We emphasize that the results of the analysis presented here have to be regarded as qualitative. In fact, the relative scarcity of currently available data does not allow to claim the detection, within reasonable accuracy, of the detailed causal structure of glacial-interglacial dynamics on the whole spectrum of time scales.
From a methodological point of view, our work clearly shows that the 
direct application of causality detection methods to  unfiltered data may provide only a part of the story. The results reported here indicate that causality analysis can be a powerful approach to study paleoclimatic signals (such as multiple isotope records from ice cores, speleothems or sediments), provided that the data set is long enough and almost-linear interactions between the relevant climatic variables can be assumed on the temporal scales of interest.

%%%%%%%%%%%%%%%%%%%%%%%%%%%%%%%%%%%%%%%%%%%%%%%%%%%%%%%%%%%%
\section*{Methods}
%%%%%%%%%%%%%%%%%%%%%%%%%%%%%%%%%%%%%%%%%%%%%%%%%%%%%%%%%%%%

%===============================================================
\subsection*{Dataset}
%===============================================================

The time series of reconstructed temperature and CO$_2$ concentration analyzed here are obtained from the EPICA Dome C ice drilling project in Antarctica, as described in Ref.~\cite{Luthi08}. Here we use the data described in Ref.~\cite{Bereiter15}, where the CO$_2$ concentration record was obtained by blending different ice cores and the chronology for the first 200 kyrs was revised and improved. In comparing the CO$_2$ concentration and temperature records, the issue of the gas-ice age difference (the so-called delta age) and its uncertainty should be kept in mind~\cite{loulergue2007new}. Here, we adopt the chronology indicated in the Ref.~\cite{Bereiter15}.
The time series of the insolation was calculated using the software provided at the site https://sites.google.com/site/geokotov/software and on the reconstructions of Ref.~\cite{Berger}.

%===============================================================
\subsection*{Response function in the presence of slow external driving}
%===============================================================
In this section we export the FDR formalism to cases in which slow time-dependent external driving is present. This is relevant in the climate context, where insolation drives the system on very slow time scales, compared to those for which experimental data are available.

First we will show that, for the considered class of models, the response function can be written in terms of correlation functions of suitably defined fast components of the dynamics. These components can be estimated, within reasonable approximations, from a proper filtering of the time series of the original variables. An example with a toy model is then discussed to illustrate the analytical results.

%=========================================================================
\subsubsection*{Application of Generalized FDR}
%=========================================================================
We will limit ourselves to the study of models in which the dynamics of the $n$ variables ${x_i},\,i=1,...,n$ representing the state coordinates can be written as
\begin{equation}
 \dot{\vect{x}}=-\mathbb{A}\vect{x}+\vect{c}f(t)+\vect{\xi}(t)\,,
\label{eq:full_dynamics}
\end{equation}
where $A$ is a $n\times n$ invertible, positive-definite and diagonalizable matrix; 
$\vect{c}$ is an $n$-dimensional vector of amplitudes, 
$\xi$ denotes a $\delta$-correlated diagonal noise $\langle \xi_i(t) \xi_j(s)\rangle = D_i\,\delta(t-s)\delta_{ij}$. 
We call $\tau_0$ the relaxation time of the free dynamics (i.e., without the forcing term), determined by the inverse of the spectral radius of $A$. The conditions on $A$ insure that the dynamics will not diverge in time. The diagonalizability requirement could actually be relaxed, as it is not essential to the proof, but it allows to simplify calculations: { see Sec.~2 of the Supplemental Material} for a brief discussion on this point.
For our application to paleoclimate time series, it is reasonable to consider a slow forcing (e.g. periodic, or quasiperidic, with long periods)
\begin{equation}
\label{eq:forzante}
 f(t)=\sum_{i=1}^l a_i \cos(t/\tau_i+\phi_i)\,,
\end{equation} 
where $\{a_i\}$ and $\{\phi_i\}$ are dimensionless constants $O(1)$. 
We assume that  the $\{\tau_i\}$ are much larger than $\tau_0$, i.e.
\begin{equation}
\tau_l\ge\tau_{l-1}\ge...\ge\tau_1\gg\tau_0\,.
\end{equation}

We are interested in the response of the system to an instantaneous perturbation $\vect{x}(0) \to \vect{x}(0) + \delta \vect{x}(0)$. In particular, we want to compute the response function~\eqref{eq:resp_obs}, assuming that we ignore the details of the model, and that we only have access to the measured trajectories of the system. The generalized FDR~\eqref{eq:resp_log} cannot be used \textit{tout court} in this context; indeed, due to the presence of the external forcing, $f(t)$, the dynamics is not Markovian, because our set of variables $\vect{x}$ does not completely describe the state of the system.

It is then useful to decompose the full dynamics \eqref{eq:full_dynamics} into ``slow'' $\vect{x}_S$ and ``fast''  $\vect{x}_F=\vect{x}-\vect{x}_S$ components, evolving as
\begin{eqnarray}
 \dot{\vect{x}}_S &=&-\mathbb{A}\vect{x}_S+ \vect{c} f(t) \label{eq:lente}
\\
\dot{\vect{x}}_F &=&-\mathbb{A}\vect{x}_F+\vect{\xi}(t)\,.
\label{eq:veloci}
\end{eqnarray} 
The above definition identifies a slow set of variables as those whose dynamics is only affected by the slow external forcing, while the uncorrelated noise is only present in the fast variables evolution.

Let us assume that the instantaneous change $\vect{x}(0) \to \vect{x}(0) + \delta \vect{x}(0)$, occurring at time $t=0$, entirely affects the fast components $\vect{x}_F$. Of course we can always make such an assumption, as the only constraint imposed by the definition~\eqref{eq:veloci} is that the sum of $\vect{x}_F(0)+\vect{x}_S(0)$ is increased by $\delta \vect{x}(0)$. Denoting with ``$\vect{x}^{(P)}$'' the perturbed dynamics  one has therefore
\begin{equation}
 \vect{x}^{(P)}(t)-\vect{x}(t)=\vect{x}^{(P)}_F(t)-\vect{x}_F(t)\quad\quad \forall t>0\,,
\end{equation} 
which follows from the independence of the evolutions, $\vect{x}_F$ and  $\vect{x}_S$. By definition, Eq.~\eqref{eq:resp_obs} implies that the response function for the complete dynamics $\vect{x}$ is equal to that of the fast variables $\vect{x}_F$.

The physical meaning of this choice is easily understood in the context of paleoclimate, where the slow dynamics can be associated to the effect of the astronomical forcing, while the behaviour of the fast components is meant to be related to the internal  climate dynamics.
%of earth. 
In this case our choice is equivalent to saying that the latter components are actually modified by an instantaneous perturbation (e.g. a large emission of CO$_2$ due to a volcano eruption), while the former, which only depend on astronomical motion, are not affected by this kind of events. 

At this point, if the trajectories of the fast components, $\vect{x}_F$,  were accessible, a plain employ of Eq.~\eqref{eq:resp_log}
would be possible, since the fast dynamics does not depend on the external forcing $f(t)$ and it is therefore Markovian. Moreover, since it is also described by a linear model, we could straightforwardly apply Eq.~\eqref{eq:respcorr} and get:
\begin{equation}
\label{eq:respcorrfast}
    \mathbb{R}(t)=\mathbb{C}_F(t)\mathbb{C}_F^{-1}(0)
\end{equation}
with
\begin{equation}
    \mathbb{C}_F(t)=\langle \vect{x}_F(t)\vect{x}^T_F(0)\rangle\,.
\end{equation}
For the class of dynamics described by Eq.~\eqref{eq:full_dynamics}, this result provides an easy way to compute the response functions, once the dynamics of the fast variables is known.

%====================================================================
\subsubsection*{Evaluating of fast correlations from data filtering}
%====================================================================
The computation of the response functions by means of Eq.~\eqref{eq:respcorrfast}  requires the evaluation of the correlation function matrix $C_F(t)$ appearing on the right hand side.
The latter is usually not accessible from experiments and observations: if a large time-scale separation is present, however, such correlation functions can be estimated by considering a proper filtering of the dynamics. 
Remarkably, the quality of the approximation increases with the separation between the time scales.
 
The idea is to replace $\vect{x}_F$ by     
\begin{equation}
\label{eq:filtrate}
\tilde{\vect{x}}(t)=\vect{x}(t)-\int_{-\infty}^{\infty}ds\,\mathcal{G}(t-s)\vect{x}(s)\,,
\end{equation}
with
\begin{equation}
 \mathcal{G}(t)=\frac{e^{-t^2/2T_w^2}}{\sqrt{2 \pi} T_w}\,,
\end{equation}
i.e. to subtract from the full dynamics a suitably defined running average. Here $T_w$ is the characteristic time-window of the Gaussian filter. 
The idea, not new~\cite{baldovin19}, is that the filtered signal mimics the behaviour of the slowly varying components, so that $\tilde{\vect{x}}(t)$ can be seen as a ``surrogate'' of $\vect{x}_F(t)$; unlike $\vect{x}_F(t)$, however, $\tilde{\vect{x}}(t)$ can be easily computed from empirical data.

One of the advantages of using a Gaussian filter\cite{filtering} relies 
on the possibility to show analytically that
\begin{equation}
  \langle \tilde{\vect{x}}(t)\tilde{\vect{x}}^T(t')\rangle\simeq\langle\vect{x}_F(t)\vect{x}_F^T(t')\rangle+ O(\max\{\tau_0/T_w,T_w^2/\tau_1^2\})\,.
\end{equation} 
The details of the proof, which involves easy but tedious computations, are reported in the Supplemental Material, { Sec.~2.} 
Here, the main point of the computation is the possibility to always find a $T_w$ such that both $\tau_0/T_w$ and $T_w^2/\tau_1^2$ are small, provided that the time-scale separation between $\tau_0$ and $\tau_1$ is large enough. 
The optimal order of magnitude for the width of the window is given by 
\begin{equation}
\label{eq:tw}
T_w \sim (\tau_0 \tau_1^2)^{1/3}\,,
\end{equation}
which ensures
\begin{equation}
 \frac{\tau_0}{T_w} \simeq \frac{T_w^2}{\tau_1^2} \ll 1\,.
\end{equation} 
Thus, the correlation functions in Eq.~\eqref{eq:respcorrfast} can be written in terms of the quantities~\eqref{eq:filtrate}, which can be straightforwardly obtained form the time series.
In other words, one has
\begin{equation}
\label{eq:respcorrfilter}
    \mathbb{R}(t) \simeq \tilde{\mathbb{C}}(t)\tilde{\mathbb{C}}^{-1}(0)
\end{equation}
with $\tilde{\mathbb{C}}(t)=\langle \tilde{\vect{x}}(t)\tilde{\vect{x}}^T(0)\rangle$.

{ Section~3.1} of the Supplemental Material contains numerical examples illustrating how the above proposed combination of filtering and generalized FDR works in practice. 
We also show, numerically, that the method is robust with respect to the presence of (small) nonlinear terms in the fast dynamics.

%====================================================================
\subsection*{Application to paleoclimate}
%====================================================================

To apply the proposed analysis to paleoclimate dynamics we assume, as a working hypothesis, that the dynamics of temperature and [CO$_2$] can be approximated by a model of the form~\eqref{eq:full_dynamics}. Here $\tau_1$ is the typical time-scale of the Milankovitch series (approximately 40 kyr), while $\tau_0$ is a characteristic time of the fast dynamics of $T$ and [CO$_2$], which we expect to be of the order of the kyr. The time-scale separation should then allow the application of our analysis, at least at 
a qualitative level.  
This scenario is supported by consistency checks which will be described in the remaining part of this Section.

%=======================================================================
\subsubsection*{Data pre-processing}
%=======================================================================
Our study of generalized FDR on paleoclimate time series has required 
a pre-processing to make the data ready for the analysis. 
First, we operated a ``data alignment'', since the time series needed to be synchronized to make the generalized FDR applicable, 
while the original data were obviously not. 
In particular, the data of temperature and [CO$_2$] were recorded on different set of times $a=\{t_1,t_2,\ldots,t_n\}$, $b=\{t'_1,t'_2,\ldots,t'_k\}$.
We used a spline interpolation method to align the two 
datasets, in such a way that the resulting series were characterised by time intervals of $0.5$ kyrs between consecutive entries (close to the original average time interval). 

A relevant point concerns the fact that temperature and CO$_2$  have different temporal resolution, with CO$_2$ showing more sparse data than temperature, { in most parts of the record}. This is reported in  Fig.~\ref{fig:spectrumclimate}(a), which displays the temporal resolution of the $T$ and [CO$_2$] records. In any case, the temporal resolution rarely becomes lower than 1 kyr, thus affecting only the shortest lag considered (0.5 kyr). {The different abundance of data between the two signals may introduce spurious statistical effects when interpolating: points generated from a set with lower density are more correlated, and this may affect the subsequent analysis, at the shortest time scales. To avoid this kind of effects we degraded the temporal resolution of the records in such a way that they were always locally comparable.}
{In particular, we divided the total observation interval (800 kyr) into 80 equal segments. For each of these 10 kyr intervals, we compared the number of available data for $T$ and [CO$_2$], and we deleted from the larger sample a number of entries equal to the difference. The data to be deleted were taken at regular intervals in the sequence.}

{The analysis of the data with undegraded temporal resolution confirms the findings reported here, and it is discussed {in Sec.~5 of the Supplemental Material}, see also Fig.~S5. { The analysis of the two halves of the signal, shown in Figs.~S6 and~S7 of the Supplementary Material }, was performed on the same temperature signal with reduced temporal resolution used here.}

After filtering, we rescaled the values of temperature and [CO$_2$] in order to be standardized: zero average and unitary variance. 
This is necessary to get rid of the degree of freedom due to the arbitrary choice of measure units, and allows to make comparisons between response functions relative to different physical quantities (see Ref.\cite{baldovin20} for a discussion on this point).

%=======================================================================
\subsubsection*{Width of the time window}
%=======================================================================

%------------------------- Fig.Spectra ------------------------------
\begin{figure}[t]
\centering
  \includegraphics[width=\linewidth]{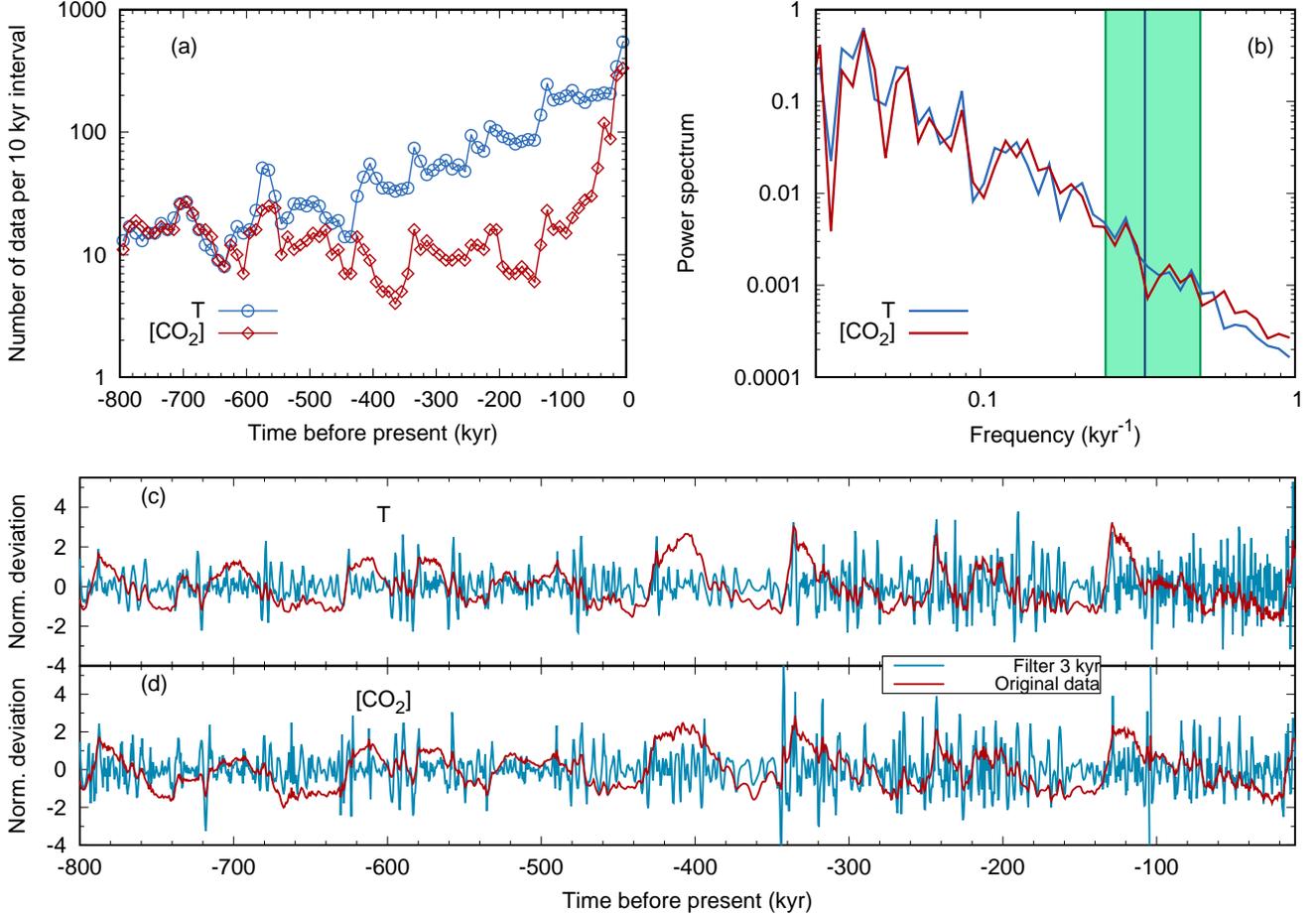}
  \caption{\label{fig:spectrumclimate} { (a) Temporal resolution of the temperature (azure circles) and CO$_2$ concentration (red diamonds) along the record. In the analysis, we have degraded the temporal resolution of temperature to match that of CO$_2$.} (b) Power spectrum of the time series for $T$ (azure) and [CO$_2$] (red). The vertical green band indicates the frequencies corresponding to the three values of $T_w$ employed in the Results section. 
 (c, d) High-pass-filtered versions of the T and [CO$_2$] signals, for the whole 800-kyr period, obtained by using a threshold $T_w =$ 3 kyr, compared to the original datasets. Data are shown as normalized deviations from average.  }
\end{figure}
%--------------------------------------------------------------------

{ The power spectra of $T$ and [CO$_2$], Fig.~\ref{fig:spectrumclimate}(b), show a common regime, smoothly decreasing with almost power-law dependence at time scales shorter than about 10 kyr. As such, there is no spectral gap at a precise frequency. We applied the high-pass filter at a time scale that is much shorter than the scale of the astronomical forcing. 
In the Results sections we take as a reference value $T_w=3$ kyr, and we repeated the analysis also for $T_w=2$ kyr and $T_w=4$ kyr. The effect of this filtering procedure on the data series can be estimated by looking at Fig.~\ref{fig:spectrumclimate}~(c) and ~(d), where the signals before and after applying the filter are plotted for the whole dataset. In general, the millennial-scale oscillations tend to be more pronounced during the glacial periods.}

%=======================================================================
\subsubsection*{Effect of the filter}
\begin{figure}[t]
\centering
  \includegraphics[width=\linewidth]{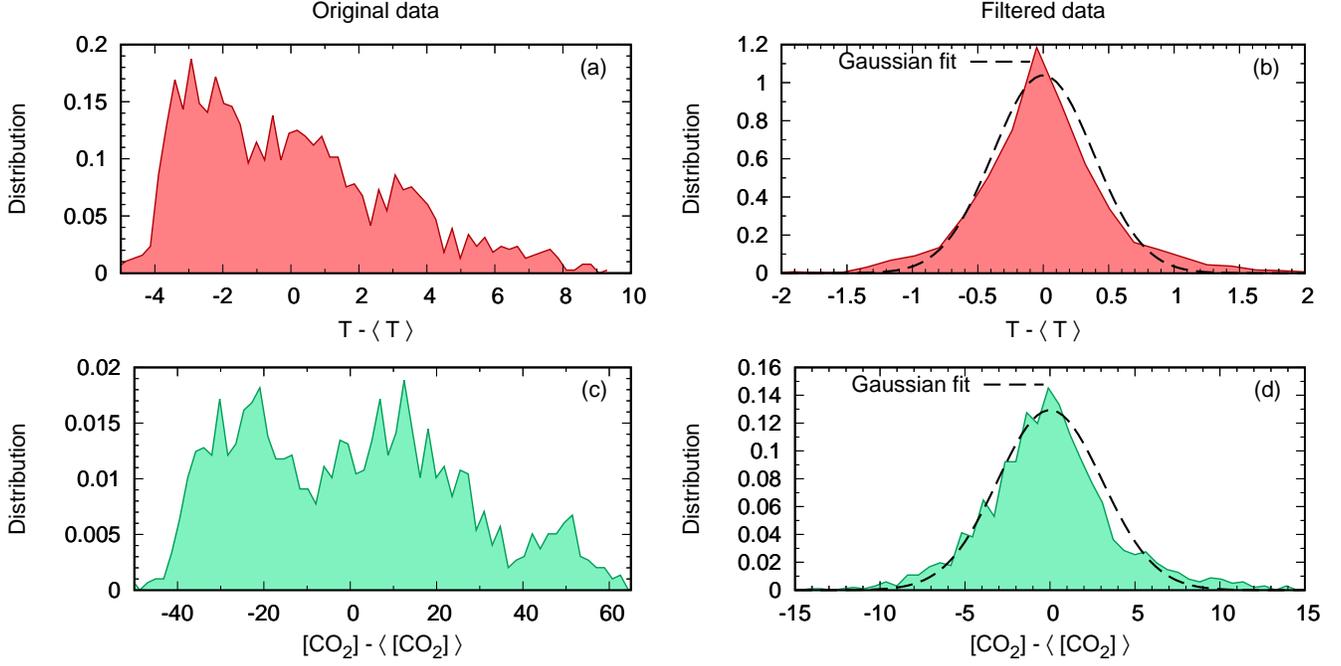}
  \caption{\label{fig:distrclimate} Histograms of the centered variables
  $T-\langle T \rangle$ and  $[\mbox{CO}_2] - \langle [\mbox{CO}_2]\rangle$
  before [Panels (a),(c)] and after the filtering [Panels (b),(d)]. 
  The filter makes the distribution of the signal Gaussian-like (compare with the Gaussian fits, dashed line in the right column); at 
  the same time it reduces the excursion of the signal, since it removes 
  the large oscillations due to the slow external forcing. Binning: for each plot, 60 bins of equal size are considered, ranging from the lowest to the highest recorded value.}
\end{figure}

%=======================================================================
The action of the Gaussian filter~\eqref{eq:filtrate} is clearly visible in Fig.~\ref{fig:distrclimate}, reporting the histograms of $T-\langle T \rangle$ and $[\mbox{CO}_2] - \langle [\mbox{CO}_2]\rangle$, before and after filtering. 
What can be deduced by the comparison between the original and the final distributions is that the filtering has a twofold effect: it makes the distributions 
Gaussian-like and, at the same time, it reduces the excursion of the signal.

\begin{figure}[h!]
\centering
  \includegraphics[width=\linewidth]{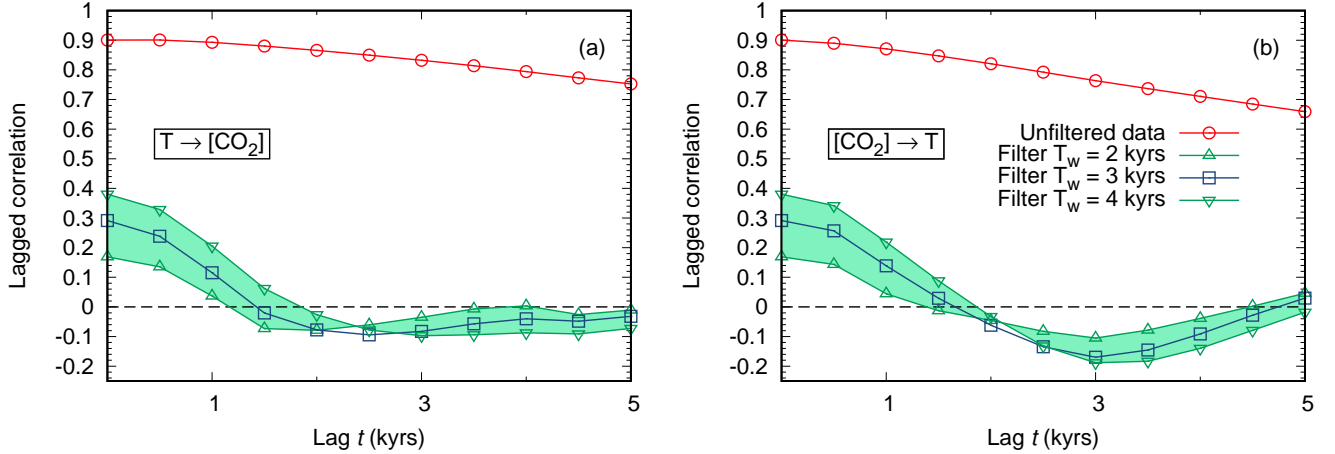}
  \caption{\label{fig:correlations} Lagged cross correlations for the T and [CO$_2$] signals, before and after applying the Gaussian filter discussed in the text.}
\end{figure}

The former fact can be seen as an hint (although not a proof) that the filtered variables have an almost-linear dynamics, which is our working hypothesis. The latter indicates that most of the variability is on longer time scales, where the effects of the slow forcing and/or of stronger nonlinear climatic responses (both being removed when the signal is filtered), are non-negligible.
%--------------------------- Histograms ---------------------------------
%-------------------------------------------------------------------
Finally, in Fig.~\ref{fig:correlations} we show the cross correlations of the two signals, before and after the filtering procedure. From these plots it is clear that the effect of the filter consists in removing from the analysis the large contributions coming from correlations on longer time scales. It should also be remarked that the cross correlations alone, even after the filter, are not very informative about the causal relations between the signals: in order to give insightful information on the causal structure of the system they must be properly combined with the self correlations, as prescribed for instance  by the generalized FDR formalism.

%--------------------------- Correlations ---------------------------------
%-------------------------------------------------------------------

\subsection*{Transfer entropy: a complementary approach}
Transfer entropy was introduced by Schreiber~\cite{shreiber00} as an indicator of the information which a given time-dependent signal $x_1(t)$ provides about a second variable $x_2(t)$. The basic idea is to measure how much information is lost about the distribution of $x_2(t)$ when the knowledge of  $x_1(t)$ is ignored.

For a two-variable Markovian system, the TE with lag $t$ is defined as
\begin{equation}
    TE_{1\to2}(t)=H[x_2|x_2](t)-H[x_2|x_1,x_2](t)
\end{equation}
where 
\begin{equation}
\label{eq:shannon}
    H[y|x](t)=-\int dx\, dy\, P(x,0;y,t) \ln P(x,0;y,t)+\int dx\,  P(x) \ln P(x)\,
\end{equation}
is the conditional Shannon entropy.
Here $P(x)$ represents the marginal $x$ probability density functions, while $P(x,0;y,t)$ is the joint distribution of $x$ at time 0 and $y$ at time $t$ assuming stationarity. If the dynamics is linear the above relations can be simplified, as shown in Ref.\cite{sun2015causal} (see 
{ Sec.~4 of} the Supplemental Material for details): in particular, it can be shown that TE is a (complicated) function of two-points correlations functions.

From the point of view of the application to paleoclimatic series, the study of TE presents therefore the same difficulties encountered in the case of generalized FDR: the system under study is not Markovian, and the limited amount of data does not enable to determine a reliable functional form for the probability density functions appearing in Eq.~\eqref{eq:shannon}.
It may thus be expected that the above-discussed filtering procedure, by isolating the fast component of the correlation functions, allows to get rid of the spurious long-time-scale correlations, as in the case of the generalized FDR. 
This expectation seems to be confirmed by Fig.~\ref{fig:fdrclimate}.

It is worth mentioning that an alternative rigorous attempt to assess causation was due to Granger\cite{granger69}, who suggested that the link $x_1\to x_2$ holds if the knowledge of the past history of $x_1$ enhances the ability to predict future values of $x_2$. Remarkably, Granger causality and TE have been shown to be equivalent in linear auto-regressive systems\cite{barnett09}.

With respect to the detection of causal links in a given system, TE and Granger's approach can be regarded as complementary to responses. While the former focuses on our ability to \textit{predict} future values of the considered process, the latter aims at defining the interaction mechanisms internal to the system\cite{baldovin20}.  

%%%%%%%%%%%%%%%%%%%%%%%%%%%%%%%%%%%%%%%%%%%%%%
\bibliography{sample}
%%%%%%%%%%%%%%%%%%%%%%%%%%%%%%%%%%%%%%%%%%%%%%

%%%%%%%%%%%%%%%%%%%%%%%%%%%%%%%%%%%%%%%%%%%%%%%
\section*{Acknowledgements}
%%%%%%%%%%%%%%%%%%%%%%%%%%%%%%%%%%%%%%%%%%%%%%%
We sincerely thank two anonymous reviewers whose insightful comments allowed us to greatly improve the presentation of the results. We are grateful to Carlo Barbante for relevant advice and for giving us important information on the data. The data used in this work have been  provided by the EPICA project team and are based on Refs.\cite{Luthi08,Bereiter15}. M.B. F.C. and A.V. acknowledge the support from the
MIUR PRIN 2017 project 201798CZLJ.

\end{document}

\preprint{}

\author{Marco Baldovin}
 \affiliation{Dipartimento di Fisica, Universit\`a ``Sapienza'', Rome  I-00185, Italy}%Lines break automatically or can be forced with \\
\affiliation{Universit\'{e}  Paris-Saclay,  CNRS,  LPTMS,  91405,  Orsay,  France}%
 \email{marco.baldovin@universite-paris-saclay.fr}
\author{Fabio Cecconi}
\affiliation{CNR-Istituto dei Sistemi Complessi and INFN, Unità di Roma 1, I-00185, Rome, Italy}%
\author{Antonello Provenzale}%
\affiliation{CNR-Istituto di Geoscienze e Georisorse, Pisa, I-56124 Italy}%
\author{Angelo Vulpiani}%
\affiliation{Dipartimento di Fisica, Universit\`a ``Sapienza'', Rome  I-00185, Italy}%

\date{September 12, 2022}% It is always \today, today,

\title{Supplemental Material}

  \newcommand{\av}[1]{\left\langle#1\right\rangle}
  \newcommand{\cbr}[1]{\left(#1\right)}
  \newcommand{\sbr}[1]{\left[#1\right]}
  \newcommand{\bbr}[1]{\left\{#1\right\}}
  
  \newcommand{\moby}{\mu}
  \newcommand{\frict}{\gamma}
  \newcommand{\mean}{\lambda}
  \newcommand{\tmean}{\widetilde{\lambda}}
  \newcommand{\stiff}{k}
  \newcommand{\tstiff}{\widetilde{k}}
  \newcommand{\stud}{k}
  \newcommand{\tstud}{\widetilde{k}}
  
  \newcommand{\dimless}[1]{{#1}^*}
  
  \newcommand{\levy}{L\'{e}vy }
  
  \newcommand{\addnew}[1]{#1}

\newcommand{\newversion}[1]{#1}

\DeclareRobustCommand{\vect}[1]{
  \ifcat#1\relax
    \boldsymbol{#1}
  \else
    \mathbf{#1}
  \fi}

\begin{abstract}
The supplemental material presented here aims at providing further information about our research, in particular about the analytical results used in the main text and an analysis of the possible issues related to the non-homogeneous density of the available data along the considered time interval. In what follows, the reader can find:
\\1) a complete derivation of the generalized Fluctuation Dissipation Relation;
\\2) the estimation of the error associated to the filtered procedure discussed in the main text;
\\3) a series of numerical examples which illustrate the method (and show its robustness against the inclusion of small nonlinear terms in the fast dynamics);
\\4) explicit formulas for the transfer entropy in the linear case;
\\5) further methodological remarks for the application of our method to paleoclimate datasets.
\end{abstract}

\flushbottom
\maketitle
\thispagestyle{empty}
\noindent

%%%%%%%%%%%%%%%%%%%%%%%%%%%%%%%%%%%%%%%%%%%%%%%%%%%%%%%%%%%%%%%%%%%%%%%%%%%
\section{Derivation of the generalized FDR}
%%%%%%%%%%%%%%%%%%%%%%%%%%%%%%%%%%%%%%%%%%%%%%%%%%%%%%%%%%%%%%%%%%%%%%%%%%%
In this section we briefly sketch the derivation of formula (2) of the main text. A more detailed exposition can be found in Ref.\cite{falcioni90}.
Let ${\bf x}(t) = [x_1(t),\ldots,x_n(t)]$ be a multivariate Markov process of dimension
$n$, 
whose stationary PDF $p_{st}({\bf x})$ is smooth and nonvanishing. 
We want to understand the effect on the dynamics ${\bf x}(t)$ of an instantaneous small perturbation   
$\vect{\epsilon} = (\epsilon_1,\ldots,\epsilon_n)$ performed at time $t=0$, by measuring the displacement it generates on the averages 
\begin{equation}
\delta \langle {\bf x}(t) \rangle\; = 
\langle {\bf x}(t) \rangle_{\epsilon} - \langle {\bf x}(t) \rangle_0\;,
\label{eq:disp}
\end{equation}
where the first term corresponds to the perturbed dynamics and the second 
to the original (unperturbed) one. 
The average should be interpreted as carried out over many realizations of the perturbed and the original dynamics.
The analytical computation of Eq.\eqref{eq:disp} requires the knowledge of the joint probabilities of ${\bf x}(t)$ and ${\bf x}_0$, that for a Markov process can be expressed as 
%\begin{eqnarray}
%P[{\bf x}(t),{\bf x}_0] &=& p_{st}({\bf x}_0)\; W_t({\bf x},t| {\bf x}_0) \\
%P_{\epsilon}[{\bf x}(t),{\bf x}_0] &=& p_{st}({\bf x}_0)\; W_t({\bf x},t| {\bf %x}_0 +\vect{\epsilon}) 
%\end{eqnarray}
\begin{align}
P[{\bf x}(t),{\bf x}_0] &= p_{st}({\bf x}_0)\; W({\bf x},t\,|\,{\bf x}_0) \\
P_{\epsilon}[{\bf x}(t),{\bf x}_0] &= p_{\epsilon}({\bf x}_0)\; W({\bf x},t\,|\, {\bf x}_0) =  p_{st}({\bf x}_0-\epsilon)\; W({\bf x},t\,|\,{\bf x}_0)
\end{align}
where as the perturbation involves only the initial condition, the 
probability density of the perturbed system is nothing but a rigid shift of the invariant distribution of the unperturbed system; for example, in the scalar case if $p_{st}(x_0) = 1 / \sqrt{2 \pi \sigma^2} \exp[-x^2/(2\sigma^2)]$ is a Gaussian with zero average, the perturbation will bring the system in a Gaussian distribution with average $\epsilon$, i.e. $ 1 / \sqrt{2 \pi \sigma^2} \exp[-(x-\epsilon)^2/(2\sigma^2)]$.
Moreover, because the perturbation affects only initial states, the evolution rule in unchanged thus both systems will share the same transition probability (propagator) $W({\bf x},t\,|\,{\bf x}_0)$, see Ref.~\cite{falcioni90}.
So we can write,
%In the spirit of the response theory the average are referred to the unperturbed system, this is the reason of the presence of $p_{st}({\bf x}_0)$ in both probabilities.
%In more physical terms, we assume that a infinitesimal, homogeneous and instantaneous perturbation $\vect{\epsilon}$ is not able to alter the stationary distribution of the system that thus coincides with the unperturbed one.
%As a consequence, the quantity \eqref{eq:disp} can be written as
%$$
%\delta \langle {\bf x}(t) \rangle = 
%\int d{\bf x}_0 \int d{\bf x}\;{\bf x}\;p_{st}({\bf x}_0)\,[W_t({\bf x},t| {\bf x}_0 +\vect{\epsilon}) - W_t({\bf x},t|{\bf x}_0)]\;.
%$$
%A simple change of variable ${\bf x}_0\to {\bf x}_0 + %\vect{\epsilon}$ allows the above expression to be recast as 
$$
\delta\langle {\bf x}(t) \rangle =
\int d{\bf x}_0 \int d{\bf x}\;  
\dfrac{p_{st}({\bf x}_0-\vect{\epsilon}) - p_{st}({\bf x}_0)}{p_{st}({\bf x}_0)}\;p_{st}({\bf x}_0)\, 
{\bf x}\,W_t({\bf x},t|{\bf x}_0)\;.
$$
Since $|\vect{\epsilon}|\ll 1$, the above expression can be expanded to the first order in $|\vect{\epsilon}|$, leading to the formula
$$
\delta \langle {\bf x}_i(t) \rangle = 
- \sum_{j=1}^{n} \epsilon_j \Bigg\langle x_i \dfrac{\partial \ln p_{st}({\bf x})}{\partial x_j}\bigg|_0 \Bigg\rangle.    
$$
Then the response function of the generalized (FDR) reads 
\begin{equation}
\label{eq:response}
 R_{ij}(t) =  
-\Bigg\langle x_i(t) \frac{\partial\ln p_{st}({\bf x})}{\partial x_j}
\Big|_{{\bf x}(0)} \Bigg\rangle\,,
\end{equation}
where the average $\langle \cdot \rangle$ is computed on the unperturbed system, and $R_t$ is the matrix of the linear response functions.
The above equation is valid if the system admits a (sufficiently smooth) invariant distribution, $p_{st}({\bf x})$.

When the response formula \eqref{eq:response} is applied to a linear Markov process, namely a Ornstein-Uhlenbeck process (OUP), we obtain exactly Eq.(3) of the main text.
The OUP is defined by the evolution
\begin{equation}
\label{eq:evo}
\dfrac{d{\bf x}}{dx}= \mathbb{A}{\bf x}_t+ \mathbb{B}\vect{\eta}(t)
\end{equation}
where $\mathbb{A}$ and $\mathbb{B}$ are constant $n\times n$ matrices, and $\vect{\eta}(t)$ is an array of $n$ independent delta-correlated Gaussian noises. 
The stationary PDF of the OUP, that is required in Eq.\eqref{eq:response}, is known to be \cite{gardiner1985handbook}
\begin{equation*}
p_s({\bf x}) = \dfrac{1}{\sqrt{\mathrm{Det}(2\pi\Sigma})} 
\exp\bigg\{-\dfrac{1}{2} {\bf x}^T \Sigma^{-1} {\bf x}\bigg\}
\end{equation*}
where $\Sigma = \mathrm{var}({\bf x}) = \langle {\bf x} {\bf x}^T \rangle$ is the stationary variance, that for the OUP is
$$
\Sigma = \int_{0}^{\infty} dt e^{-\mathbb{A}t}\mathbb{B}\mathbb{B}^{T}e^{-\mathbb{A}^{T}t},
$$
thus independent of time.
Now by applying Eq.\eqref{eq:response}, we straightforwardly obtain  
\begin{equation}
\mathbb{R}(t) = \mathbb{C}(t) \Sigma^{-1}  
\label{eq:respcorr}
\end{equation}
that is exactly the formula (3) of the main text, upon realizing that 
$\Sigma = \mathbb{C}(0)$.

Eq.~\eqref{eq:response}  expresses  the  link among responses and correlators of every orders, it allows the response functions to be computed from suitable correlation functions, provided that the functional form of $P_s({\bf  x})$ is known either a priori or inferred from data (often a completely non-trivial task).

%==================================================================
\section{Evaluation of the error}
%==================================================================
In this section we provide an esteem of the error associated with the filtering procedure discussed in the main text. In short, we compute the correlation functions of the filtered process showing that they are equal to those of the fast variables defined in the Methods section, but for an additive constant which depends on the time scale separation between the fast dynamics, the slow dynamics and the filtering window $T_w$.

We start by noticing that Eq.~(12) of the main text implies
\begin{equation}
\label{eq:supp1}
 \tilde{\vect{x}}(t)=\vect{x}_S(t)+\vect{x}_F(t)-\int_{-\infty}^{\infty}ds\,\mathcal{G}(t-s)\vect{x}_S(s)-\int_{-\infty}^{\infty}ds\,\mathcal{G}(t-s)\vect{x}_F(s)\,.
\end{equation} 
The slow-variable evolution is ruled by Eq.~(7) of the main text, whose complete solution is
\begin{equation}
 \vect{x}_S(t)=e^{-\mathbb{A}t} \vect{x}_S(0)+\int_0^{t}ds\, e^{-\mathbb{A}(t-s)}\vect{c} f(s)\,.
\end{equation} 
The first term on the right hand side is a transient, whose contribution becomes negligible as soon as $t$ is larger than a few $\tau_0$. To evaluate the integral, let us remember that $f(t)$ is a slow-varying function, so that $f(s)\simeq f(t)-f'(t) (t-s)$ with $f'(t)\simeq O(\tau_1^{-1})$. We get therefore 
\begin{equation}
\label{eq:lenteapprox}
 \vect{x}_S(t)\simeq \mathbb{A}^{-1}\vect{c} f(t)+O(\tau_0/\tau_1)\,;
\end{equation} 
as a consequence, the third term of the right hand side of Eq.~\eqref{eq:supp1} reads
\begin{equation}
\begin{aligned}
 \int_{-\infty}^{\infty}ds\,\mathcal{G}(t-s)\vect{x}_S(s)&\simeq \mathbb{A}^{-1}\vect{c} \sum_{j=1}^l a_j \int_{-\infty}^{\infty}ds\,\mathcal{G}(t-s)\cos(s/\tau_j+\phi_j)+O(\tau_0/\tau_1)\\
 &\simeq \mathbb{A}^{-1}\vect{c} \sum_{j=1}^l a_j \cos(t/\tau_j+\phi_j)e^{-T_w^2/2\tau_j^2}+O(\tau_0/\tau_1)\\
 &\simeq \vect{x}_S + O(\max\{T_w^2/\tau_1^2,\tau_0/\tau_1\})\,,
\end{aligned}
\end{equation} 
where also Eq.~(5) of the main text has been exploited.
The filtered variables can thus be approximated as
\begin{equation}
  \tilde{\vect{x}}(t)\simeq\vect{x}_F(t)-\int_{-\infty}^{\infty}ds\,\mathcal{G}(t-s)\vect{x}_F(s) + O(\max\{T_w^2/\tau_1^2,\tau_0/\tau_1\})\,.
\end{equation} 
We want to estimate the correlation functions appearing in the generalized FDR for linear dynamics. In the light of the above, one has
\begin{equation}
 \langle \tilde{\vect{x}}(t)\tilde{\vect{x}}^T(t')\rangle\simeq \left\langle \left( \vect{x}_F(t)-\int_{-\infty}^{\infty}ds\,\mathcal{G}(t-s)\vect{x}_F(s) \right)  \left( \vect{x}_F(t')-\int_{-\infty}^{\infty}ds\,\mathcal{G}(t'-s)\vect{x}_F(s) \right)^T \right\rangle + O(\max\{T_w^2/\tau_1^2,\tau_0/\tau_1\})\,.
\end{equation} 
The product in the average in the r.h.s. leads to four terms, one of which is $\langle\vect{x}_F(t)\vect{x}^T_F(t')\rangle$. We have to show that the remaining ones are negligible.

To estimate these terms it is important to remind the properties of the matrix $\mathbb{A}$ and the fact that the dynamics $\vect{x}_F$ is given by Eq.~(8) of the main text, so that\cite{gardiner1985handbook}
\begin{equation}
 \langle\vect{x}_F(t)\vect{x}_F^T(t')\rangle=\begin{cases}
   e^{-\mathbb{A}(t-t')}\Sigma      \quad &\mbox{if}\quad   t> t'\\
   \Sigma e^{-\mathbb{A}^T(t'-t)}      \quad &\mbox{if} \quad  t< t'
                               \end{cases}
\end{equation} 
where $\Sigma$ is the covariance matrix $\Sigma=\langle\vect{x}_F(t)\vect{x}_F^T(t)\rangle$. 
\begin{equation}
\begin{aligned}
 \left\langle  \vect{x}_F(t) \int_{-\infty}^{\infty}ds\,\mathcal{G}(t'-s)\vect{x}^T_F(s) \right\rangle=&\int_{-\infty}^{t}ds\,\mathcal{G}(t'-s)e^{-\mathbb{A}(t-s)}\Sigma +\Sigma\int_{t}^{\infty}ds\,\mathcal{G}(t'-s)e^{-\mathbb{A}^T(t'-t)}\\
 =&\mathbb{V}^{-1}\int_{-\infty}^{t}ds\,\mathcal{G}(t'-s)e^{-\mathbb{A}_D(t-s)}\mathbb{V}\Sigma +\Sigma \mathbb{U}^{-1}\int_{t}^{\infty}ds\,\mathcal{G}(t'-s)e^{-\mathbb{A}_D^T(t'-t)}U
\end{aligned}
\end{equation} 
where
\begin{itemize}
 \item $\mathbb{V}$ is the matrix that diagonalizes $\mathbb{A}$;
 \item $\mathbb{U}=(\mathbb{V}^T)^{-1}$ is the matrix diagonalizing $\mathbb{A}^T$;
 \item $\mathbb{A}_D$ is the corresponding diagonal matrix: $\mathbb{A}=\mathbb{V}^{-1}\mathbb{A_D}\mathbb{V}\,$, $\mathbb{A}^T=\mathbb{U}^{-1}\mathbb{A}_DU$.
\end{itemize}

In the above reasoning we have exploited our hypothesis about the diagonalizability of the matrix $A$. This condition simplifies the computation, as it allows to write the exponential of the matrix in a simple way. Let us stress, however, that a similar calculation could be also carried out for generic invertible matrixes, by making use of the Jordan normal form to compute the exponentials.
The above integrals are now diagonal matrices,  and we can evaluate the $j$-th diagonal element as
\begin{subequations}
\begin{equation}
    \int_{-\infty}^{t}ds\,\mathcal{G}(t'-s)e^{-a^j_D(t-s)}=\frac{1}{2}e^{a^j_D(t'-t)+(a^j_D)^2T_w^2/2}\text{erfc}\left[\frac{t'-t+a^j_DT_w^2}{\sqrt{2}T_w}\right]\simeq \frac{e^{-(t'-t)^2/2T_w^2}}{\sqrt{2\pi}a^j_DT_w}
\end{equation}
\begin{equation}
    \int_{t}^{\infty}ds\,\mathcal{G}(t'-s)e^{-a^j_D(t'-t)}=\frac{1}{2}e^{a^j_D(t-t')+(a^j_D)^2T_w^2/2}\text{erfc}\left[\frac{t-t'+a^j_DT_w^2}{\sqrt{2}T_w}\right]\simeq \frac{e^{-(t'-t)^2/2T_w^2}}{\sqrt{2\pi}a^j_DT_w}
\end{equation}
\end{subequations}
where $a_D^j$ is the  $j$-th diagonal element of the matrix $A_D$, and we have exploited the asymptotic expansion $\text{erfc}(x)\simeq e^{-x^2}/(x\sqrt{\pi})$, valid for $x\gg1$. 
Since we are indeed interested in the time-range in which $t-t'$ is at most $O(T_w)$, and recalling that the spectral radius of $A$ is order $\tau_0^{-1}$, we can conclude that the above terms are at most $O(\tau_0/T_w)$. As a consequence,
\begin{equation}
 \left\langle  \vect{x}_F(t) \int_{-\infty}^{\infty}ds\,\mathcal{G}(t'-s)\vect{x}^T_F(s) \right\rangle \simeq O(\tau_0/T_w)\,.
\end{equation} 
Reasoning in the same way for the remaining terms one obtains
\begin{equation}
  \langle \tilde{\vect{x}}(t)\tilde{\vect{x}}^T(t')\rangle\simeq\langle\vect{x}_F(t)\vect{x}_F^T(t')\rangle+ O(\max\{\tau_0/T_w,T_w^2/\tau_0^2\})\,,
\end{equation} 
since terms $O(\tau_0/\tau_1)$ are always negligible w.r.t. $O(\tau_0/T_w)$.

The sources of error in the approximation are therefore two. To minimize the error, one has to choose $T_w$ of the order of $T_w^{\star}=(\tau_0 \tau_1^2)^{1/3}$.

%==================================================================
\section{Numerical examples}
%==================================================================
To illustrate how the above proposed combination of filtering and 
GFDR works in practice, we consider some pedagogical examples based on suitably defined toy models. 

%=======================================================================
\subsection{A toy model with linear fast dynamics}
%=======================================================================
Let $\vect{x}$ be the two-dimensional system defined by Eq.~(4) of the main text, where $A$ is the $2\times2$ matrix 
\begin{equation}
\label{eq:linearmodel}
\mathbb{A}=\frac{1}{\tau_0}\begin{pmatrix}
   1 && a_{12}\\
   a_{21} && 1  
  \end{pmatrix},
\end{equation}
and the forcing reads
\begin{equation}
\label{eq:cos}
 f(t)=\cos \left( \frac{2\pi t}{\tau_1(1+\varepsilon)}\right)+\cos \left( \frac{2\pi t}{\tau_1(1-\varepsilon)}\right)\,.
\end{equation} 
Finally, $\vect{c}=(1/2,1/2)$ and $D=1/\tau_0$. In this case the fastest characteristic time of the slow dynamics is $O(\tau_1)$, while the typical time-scale of the fast one is $\tau_0$.
%-------------------------------- Fig.2 -----------------------------------
\begin{figure}[ht]
\centering
  \includegraphics[width=.8\linewidth]{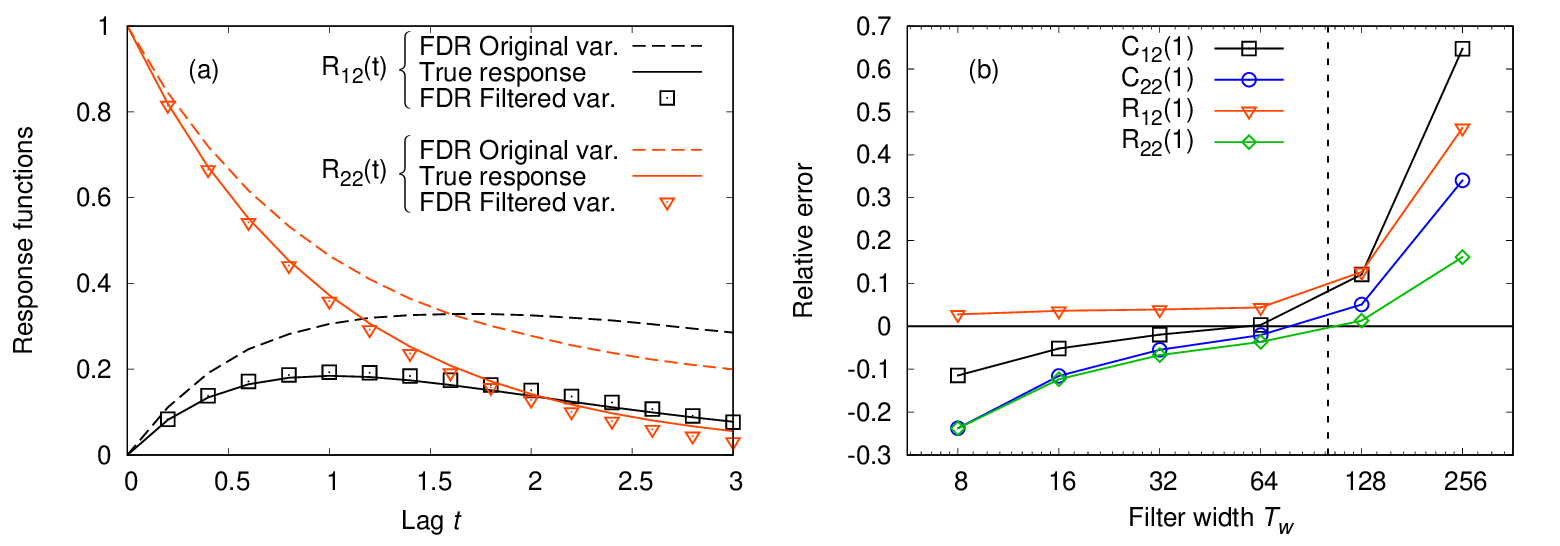}
  \caption{\label{fig:cos} Toy model with forcing, Eq.~\eqref{eq:cos}. 
  Panel (a) shows two response functions, measured from numerical simulations (solid lines), compared to the GFDR obtained with the filtered variables ($T_w=64$, points) and with the original variables (dashed lines). 
  Panel (b) shows the relative error of correlations (with respect to the fast variables) and FDR (with respect to responses) for several values of the filter constant $T_w$. The scale of the optimal value (as deduced from the theoretical argument) $T_w^{\star}=(\tau_0 \tau_1^2)^{1/3}$ is marked by a dashed vertical line.
    Here $a_{12}=-0.5$, $a_{21}=-0.05$, $\varepsilon=0.02$, $\tau_0=1$ and $\tau_1=1024$. }
\end{figure}
%--------------------------------------------------------------------------

 The above dynamics can be easily simulated with the standard Euler-Maruyama algorithm \cite{kloeden1992stochastic}, and the proposed analysis can be applied to the numerically generated trajectories. 
 The outcomes are shown in Figure~\ref{fig:cos}. In panel (a) the solid lines represent the actual response functions, computed according to the definition, as in Eq.~(1) of the main text (an average over a large number of realizations is considered). The dashed lines are obtained by naively applying the formula of generalized FDR valid for linear systems, Eq.~(3) of the main text, to the dynamics of $\vect{x}$, which is neither Markovian nor linear. Finally, points are obtained by applying the generalized FDR to the properly filtered variables. This latter procedure is in good agreement with the results obtained by direct measure of response, as predicted by our analytical argument., while the naive analysis without filtering leads to quite misleading results.
 Panel (b) shows the relative error of correlation functions and GFDR as a function of $T_w$, the width of the Gaussian filter. As expected, the best results are obtained for $T_w$ close to $(\tau_0 \tau_1^2)^{1/3}$.

%==========================================================================
\subsection{Inclusion of nonlinear terms}
\begin{figure}[h!]
\centering
  \includegraphics[width=.8\linewidth]{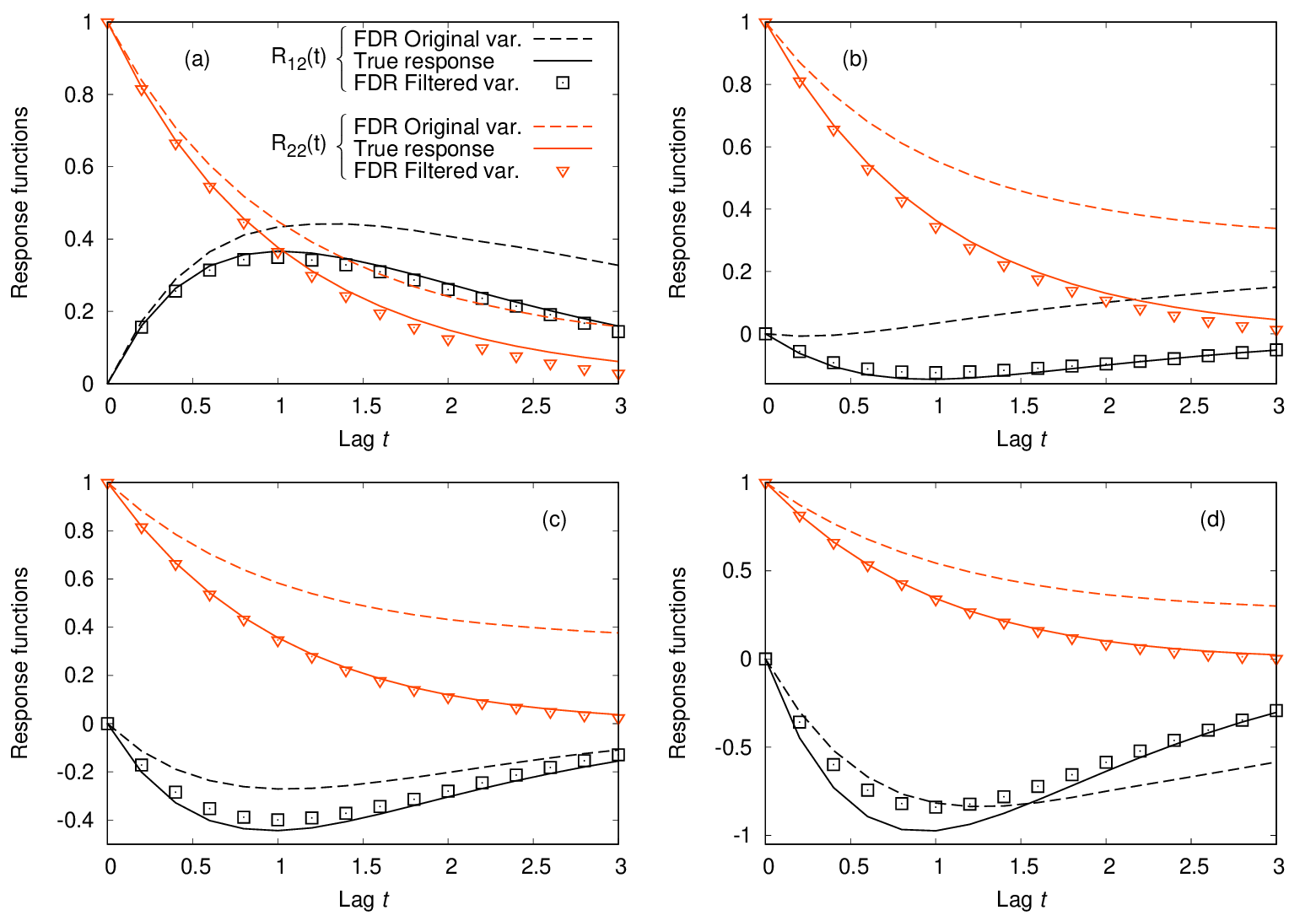}
   \caption{\label{nonl} Same as Fig.~\ref{fig:cos}(a), for model~\eqref{eq:nonlinearmodel} with different values of the coupling constant $\alpha=0.1,\,0.2,\,0.4,\,0.8$. Other parameters as in Fig.~\ref{fig:cos}.}
\end{figure}

%==========================================================================
The above argument is valid when the interactions between the variables are linear; however one may study what happens when a nonlinear interaction term is added. It is indeed known that FDR is quite robust with respect to the addition of small nonlinear perturbations of the dynamics.
In Fig.~\ref{nonl} we study model~\eqref{eq:linearmodel} where
\begin{equation}
\label{eq:nonlinearmodel}
 \dot{x}_1=A_{11}x_1+A_{12}x_2\underline{- ax_2^3} +c_1f_1(t)+ \xi_1\,,
\end{equation} 
i.e. with the addition of the underlined nonlinear term.

As the figure shows, the presence of such nonlinear perturbation does not alter in a relevant way the ability of the filtering method to reproduce the actual response from long time series of data.

%=======================================================================
\subsection{Lorenz '63 forcing}
\begin{figure}[h!]
\centering
  \includegraphics[width=.8\linewidth]{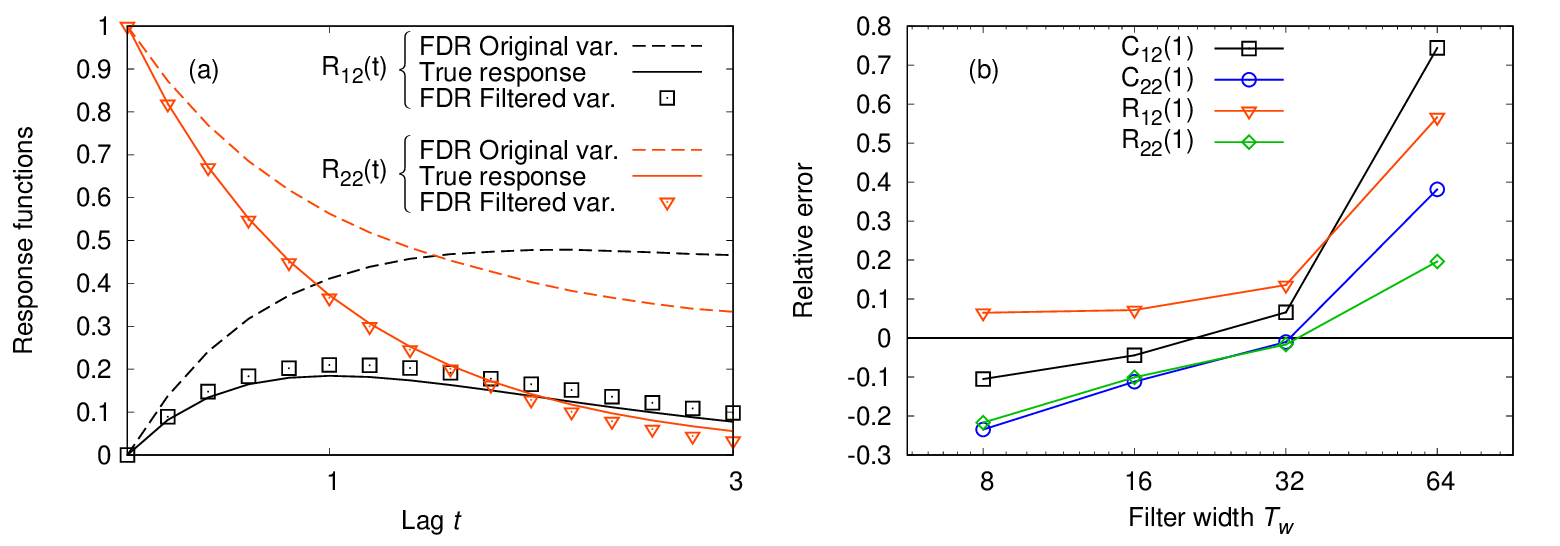}
   \caption{\label{lor} Same as Fig.~\ref{fig:cos}, with a forcing described by Eq.~\eqref{eq:lorenz}. In panel (a) the chosen value for the filter is $T=32$. Other paramteres as in Fig.~\ref{fig:cos}.}
\end{figure}

%=======================================================================
A different kind of test consists in modifying the forcing with some less regular function. To this end, we use the first component of the Lorenz '63 model
\begin{equation}
 \label{eq:lorenz}
 f(t)=x_L(t)\quad \quad  \text{where }\quad \quad
 \begin{cases}
 \dot{x}_L=\frac{1}{\tau_1}\left[ \sigma(y_L-x_L)\right]\\
 \dot{y}_L=\frac{1}{\tau_1}\left[ x_L(\rho-z_L)-y_L\right]\\
 \dot{z}_L=\frac{1}{\tau_1}\left[ x_Ly_L-\beta z_L\right]                                       
 \end{cases}
\end{equation} 
with the usual choice of the variables $\sigma=10,\, \beta=8/3,\, \rho=28$. The results are shown in Fig.~\ref{lor}.

%==================================================================
\section{Transfer entropy: explicit formulation in the linear case}
%==================================================================
The definition of transfer entropy given in the main text [Eq.~(18)] can be explicitely evaluated for the cases with linear dynamics\cite{sun2015causal,sarra2021}. Let us consider a linear $N$-dimensional Markov system $\mathbf{x}(t)=\{x_1(t),...,x_N(t)\}$ with Gaussian statistics. The Shannon entropy associated to the corresponding stationary Gaussian distribution with covariance matrix $\Sigma_\mathbf{x}$ is, but for an irrelevant additive constant,
\begin{equation}
    H_{\mathbf{x}}=\frac{1}{2}\ln |\Sigma_\mathbf{x}|\,,
\end{equation}
where $|M|$ stands for the determinant of the matrix $M$.
It is also useful to remember that the covariance matrix of a conditioned Gaussian distribution verifies
\begin{equation}
\label{eq:condgauss}
    \Sigma_{\mathbf{x}|\mathbf{y}}=\Sigma_{\mathbf{x}, \mathbf{y}}\Sigma_{\mathbf{y}}^{-1}\Sigma_{\mathbf{x}, \mathbf{y}}^T\,,
\end{equation}
where $\Sigma_{\mathbf{x}, \mathbf{y}}$ is the covariance matrix of the joint distribution.
Keeping this in mind, for the linear cases one can rewrite Eq.~(18) of the main text as
\begin{equation}
    TE_{1\to 2}(t)=\frac{1}{2}\ln \left(\frac{|\Sigma_{x_2(t)|x_2(0)}|}{|\Sigma_{x_2(t)|x_1(0),x_2(0)}|}\right)\,.
\end{equation}
Taking into account Eq.~\eqref{eq:condgauss}, the above expression can be reduced into
\begin{equation}
    TE_{1\to 2}(t)=\frac{1}{2}\ln \left( 1-\frac{\alpha_{21}(t)}{\alpha_{21}(t)-\beta_{21}(t)}\right)\,,
\end{equation}
with
\begin{subequations}
\begin{equation}
    \alpha_{21}(t)=\left[\Sigma_{22}C_{21}(t)-C_{22}(t)\Sigma_{21} \right]^2
\end{equation}
\begin{equation}
    \beta_{21}=[\Sigma_{22}^2-C_{22}(t)^2](\Sigma_{22}\Sigma_{11}-\Sigma_{21}^2)\,.
\end{equation}
\end{subequations}

%==================================================================
\section{Further methodological remarks}
%==================================================================
To apply the proposed method to real situations, it is required that synchronised and equally time-spaced data are available, so that lagged cross-correlations can be suitably computed. To this aim, the original data series have to be interpolated, as discussed in the Method section of the main text. Of course this operation may introduce subtle sources of error. For instance, if the frequency of the interpolated data is larger than the original one, spurious correlations may be introduced, basically due to repeated use of the same data to generate the new points. It is thus important to carefully check that the interpolation procedure does not alter the results.

\begin{figure}[h!]
\centering
  \includegraphics[width=.8\linewidth]{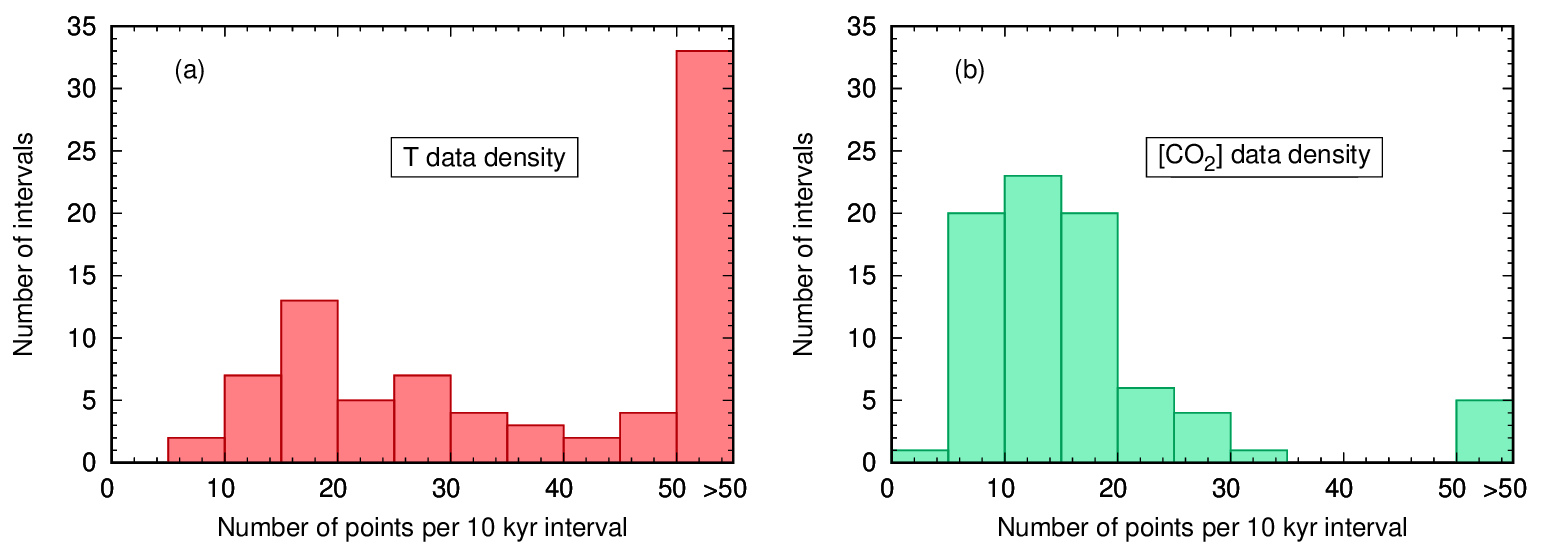}
   \caption{\label{fig:distr} Density of paleoclimate data in the considered time series. For both $T$ [panel (a)] and $[CO_2]$ [panel (b)] the dataset has been divided into 80 time intervals of 10 kyr duration, and the number of data points falling in each interval has been measured. The distribution of the data population indicates in this case that interpolations made with a frequency of 1 point per kyr, or less, are reliable since the number of intervals with lower data densities is relatively small for both samples.}
\end{figure}

%----------------------------------------------------------------------
\begin{figure}[]
\centering
  \includegraphics[width=.8\linewidth]{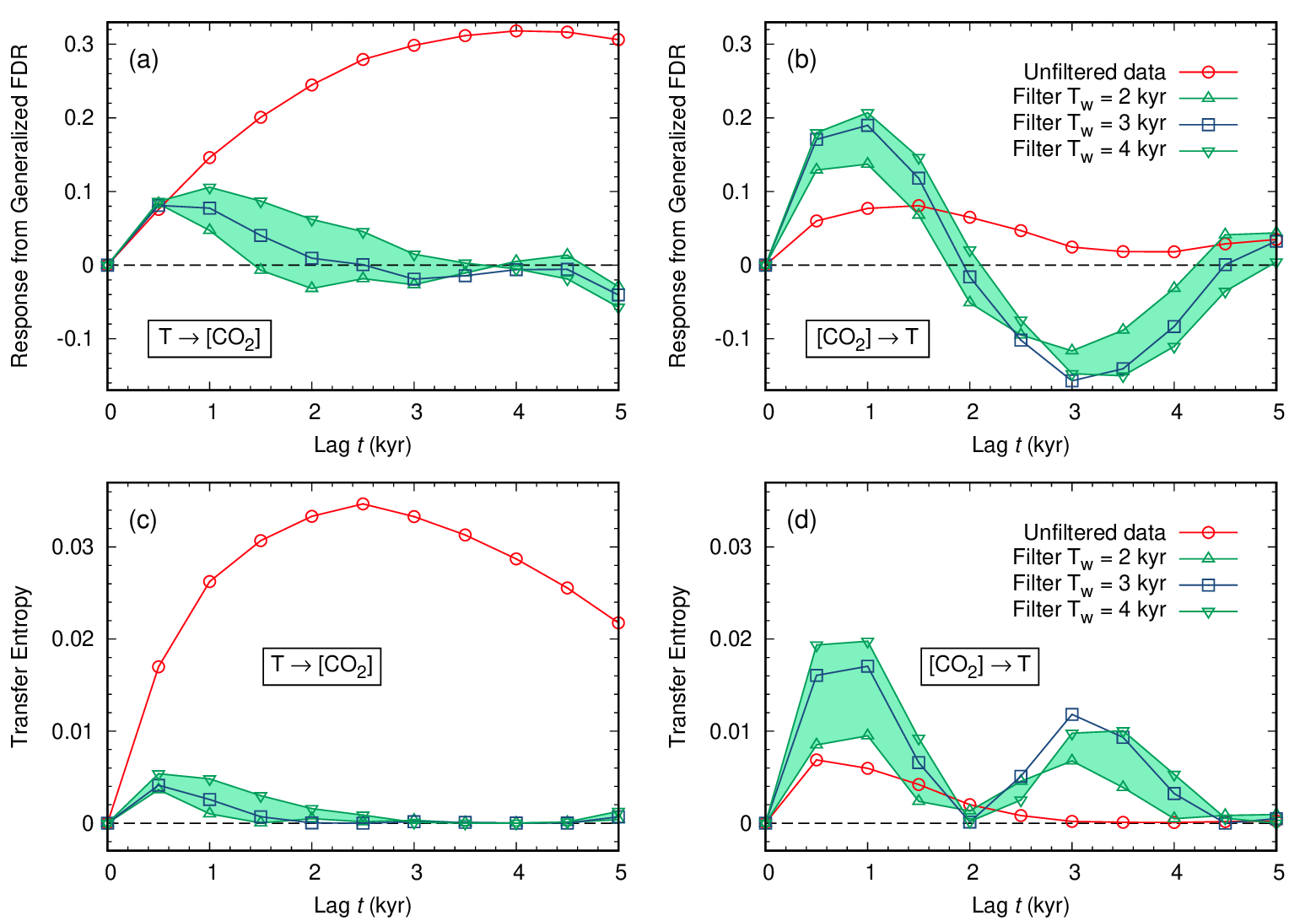}
  \caption{\label{fig:undegraded} 
  { Analysis of the mutual influence between 
  temperature $T$ and CO$_2$ concentration using the data with undegraded temporal resolution. 
  Panels (a) and (b) show the response function, computed according to the Generalized FDR. Panel (a) refers to the effect of T on  [CO$_2$], while panel (b) shows the opposite relation. Red circles represent the results of a direct application of Eq.~(3) of the main text to raw data,  apparently suggesting that $T\to$[CO$_2$] is much stronger than [CO$_2$]$\to T$. 
  The response computed from data filtered by a $T_w=3$ kyr window (blue squares) instead indicates that the impact of [CO$_2$] on $T$ becomes larger. % and it vanishes within about $2$ kyr.
  The result is robust with respect to $T_w$ variations by one kyr (green up/down triangles). A similar analysis, where TE is computed instead of the generalised FDR, is shown in Panels (c) and (d). Here, we have considered the data with their original temporal resolution, which is different for temperature and CO$_2$ concentration.} 
}
\end{figure}
%----------------------------------------------------------------------

{ Figure 2 in the Methods section already shows that the temporal resolution of the temperature and CO$_2$ records is different and varies with time. Another way to further verify whether} a non-homogeneously time-spaced dataset can be approximated by an interpolated series with a given lag relies on the analysis of the data population along the sample. The idea is to divide the total time interval into smaller segments and to count the number of data falling in each of them. The statistics of the data populations allows to determine how reliable the interpolation is, for a given value of the lag. If, for instance, most segments show a density of points which is lower than the inverse of the lag, this clearly means that in many parts of the dataset we are creating more points than the original ones. Note that this criterion is much stricter than simply considering the average lag between two data in the original series, as it takes into account the possibility of large fluctuations of the data density along the time series.

A graphical illustration of this analysis is shown in Fig.~\ref{fig:distr}. From the figure it is clear that the $[CO_2]$ dataset is the most critical, as the amount of segments containing only few points is larger. To be more quantitative, about 80\% of the considered intervals are populated by less than 20 points, meaning that the results obtained with a lag of 0.5 kyr (the first non-zero lag value in our plots) may be affected by the spurious correlations produced by the interpolation, as stated in the Method section of the main text. On the other hand, only 25\% of the intervals contain less than 10 data points, meaning that the results at 1 kyr lag and larger are quite reliable under this respect.

{ To avoid potential issues emerging from the different resolution of the temperature and CO$_2$ signals, in the analysis we opted for degrading the resolution of the temperature signal to that of the CO$_2$ concentration  {(and vice-versa, in the few intervals where the temporal resolution of [CO$_2$] is larger than that of $T$)}, see Fig.~1 in the main text. Here, in Fig.~\ref{fig:undegraded} we show the results that would be obtained without such degradation of the resolution. { In such case, the effect of [CO$_2$] on temperature becomes almost twice that of the opposite link, indicating that the estimated strength of the causal links can depend on many details, including differences in resolution. Clearly, such issues are amplified by the limited number of points available for the analysis. In any case, the main message of these results - that is, mutual causal relationships between CO$_2$ concentration and temperature, which would be lost without high-pass filtering - does not change.}}

\begin{figure}[h!]
\centering
  \includegraphics[width=.8\linewidth]{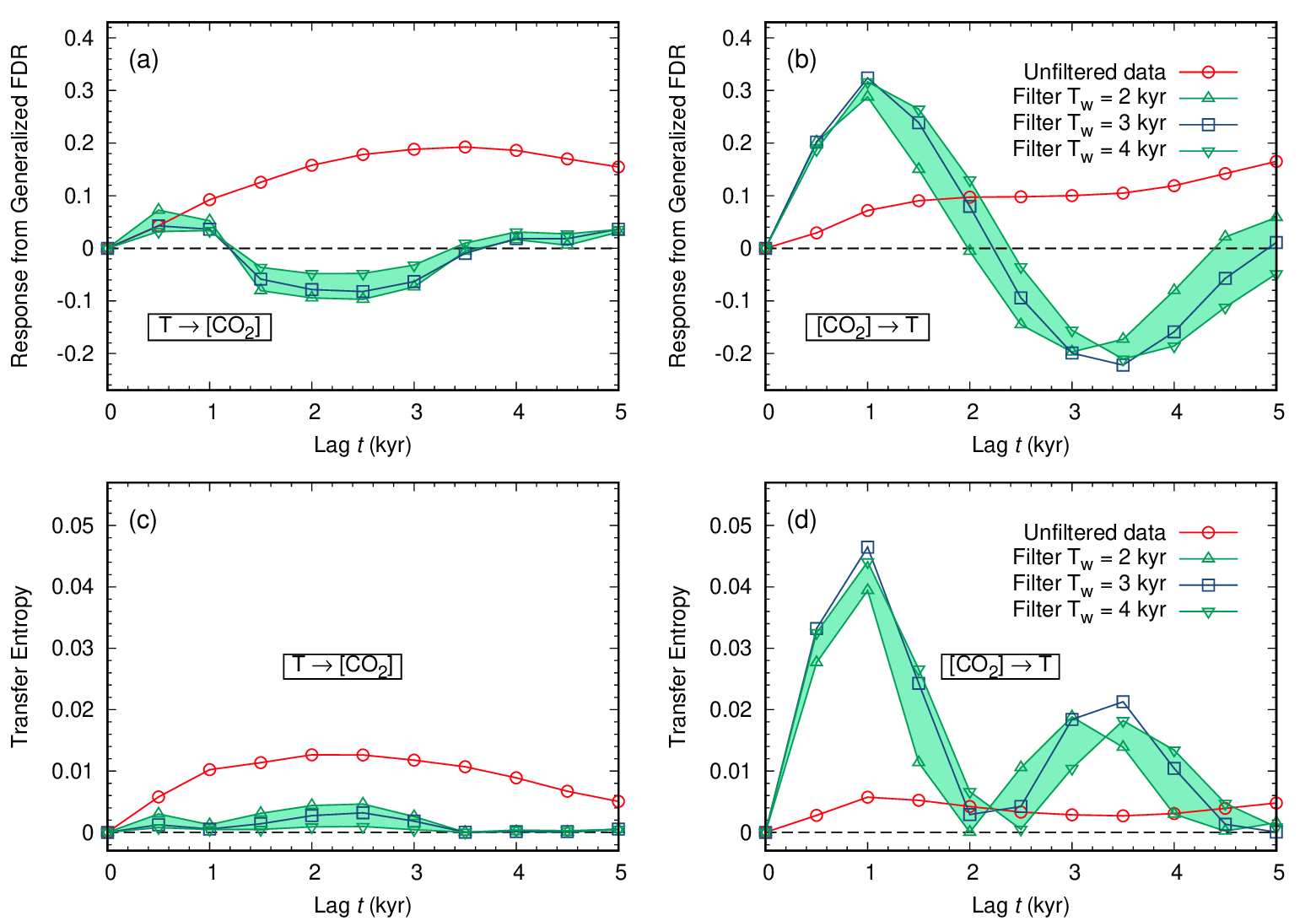}
   \caption{\label{fig:first} Analysis of the causal relations between $T$ and $[CO_2]$ from paleoclimate data, using a spline interpolation and  considering only the first 400 kyr of the record. As before, the cross terms of the generalized FDR [panels (a) and (b)], as well as the corresponding transfer entropies [panels (c) and (d)] are shown. Each quantity has been computed before and after the high-pass filtering procedure of the data series.}
\end{figure}

\begin{figure}[h!]
\centering
  \includegraphics[width=.8\linewidth]{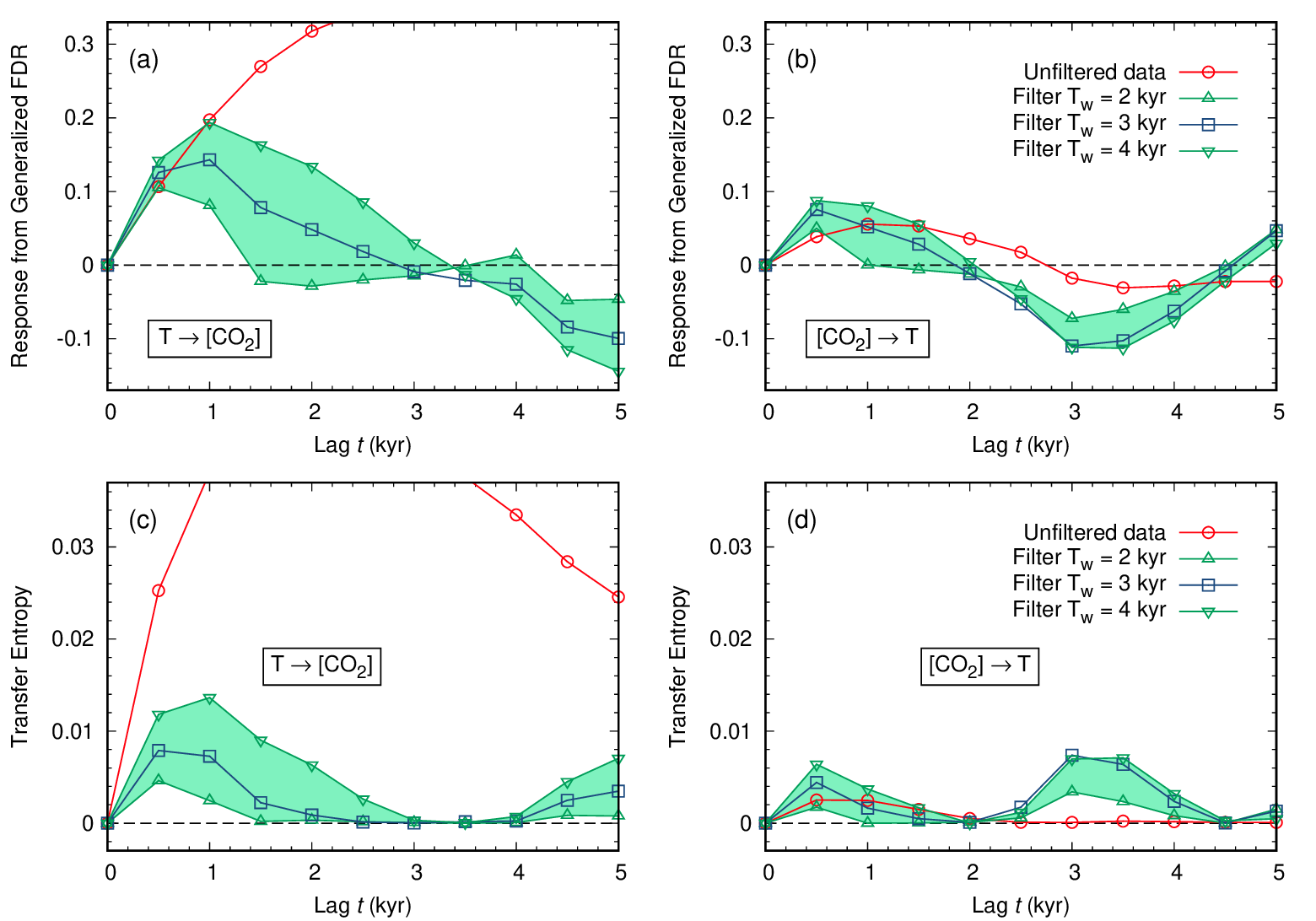}
   \caption{\label{fig:half} Analysis of the causal relations between $T$ and $[CO_2]$ from paleoclimate data, using a spline interpolation and  considering only the last 400 kyr. As before, the cross terms of the generalized FDR [panels (a) and (b)], as well as the corresponding transfer entropies [panels (c) and (d)] are shown. Each quantity has been computed before and after the high-pass filtering procedure of the data series.}
\end{figure}

An interesting test is to repeat the analysis by considering only half of the dataset, to verify whether there is a temporal dependence of the causal relationships between temperature and CO$_2$ concentration. Here, we consider either the first or the second half of the record, namely the data referring to either the first or the last 400 kyr. Of course, in this way the statistical uncertainty increases, as we are considering a more limited amount of data. 
{ 
Also in these cases, 
the high-pass filter allows to detect causal links on the scale of 1 kyr, which are hidden by the slow dynamics when performing analyses on unfiltered data. After applying the filtering procedure, the [CO$_2] \to T$ causal link appears to dominate in the first 400 kyr, as shown in Fig.~\ref{fig:first}, while it becomes weaker than the reverse link in the last 400 ky, as shown in Fig.~\ref{fig:half}.
This result can just be a statistical fluctuation generated by the limited amount of data, or it could point to a role of the temporal resolution in modulating the causal relationships of CO$_2$ and $T$, or else it could reflect a true temporal variability in the strength of the causal links between the two variables. In any case, all these results confirm the presence of millenial-scale mutual causal relationships between CO$_2$ and temperature, which could not be detected in the unfiltered record.  }

%%%%%%%%%%%%%%%%%%%%%%%%%%%%%%%%%%%%%%%%%%%%%%
\bibliography{sample}
%%%%%%%%%%%%%%%%%%%%%%%%%%%%%%%%%%%%%%%%%%%%%%